\documentclass[11pt,a4paper]{article}
\pdfoutput=1

\usepackage{amsmath}
\usepackage{amssymb}
\usepackage{verbatim}
\usepackage{bm}
\usepackage{dsfont}
\usepackage[footnotesize]{caption}
\usepackage{subfigure}
\usepackage{cite}
\usepackage{textcomp}
\usepackage{calc}
\usepackage{geometry}
\usepackage{bbm}
\usepackage{xcolor} 
\usepackage{multirow}
\usepackage{color}
\usepackage{float}
\usepackage{bbold}
\usepackage{graphicx}

\usepackage{hyperref}
\hypersetup{colorlinks=true,urlcolor=blue,linkcolor=red,citecolor=green!60!black}

\newcommand{\specialcell}[2][c]{\begin{tabular}[#1]{@{}c@{}}#2\end{tabular}}

\begin{document}


\noindent November 2016
 
\vskip 2cm

\begin{center}
{\LARGE\bf Perturbed Yukawa Textures\\\smallskip in the Minimal Seesaw Model}

\vskip 1cm

\renewcommand*{\thefootnote}{\fnsymbol{footnote}}

{\large
Thomas~Rink and Kai~Schmitz%
\,\footnote{thomas.rink / kai.schmitz \,@\, mpi-hd.mpg.de}}\\[3mm]
{\it{Max Planck Institute for Nuclear Physics (MPIK), 69117 Heidelberg, Germany}}

\end{center}

\vskip 1cm

\renewcommand*{\thefootnote}{\arabic{footnote}}
\setcounter{footnote}{0}


\begin{abstract}


\noindent We revisit the \textit{minimal seesaw model}, i.e.,
the type-I seesaw mechanism involving only two right-handed neutrinos.
This model represents an important minimal benchmark scenario for
future experimental updates on neutrino oscillations.
It features four real parameters that cannot be fixed by
the current data: two $CP$-violating phases, $\delta$ and $\sigma$,
as well as one complex parameter, $z$, that is experimentally
inaccessible at low energies.
The parameter $z$ controls the structure of the neutrino Yukawa matrix at
high energies, which is why it may be regarded as a label or index for all
UV completions of the minimal seesaw model.
The fact that $z$ encompasses only two real degrees of freedom allows us
to systematically scan the minimal seesaw model over all of its possible UV completions.
In doing so, we address the following question: Suppose $\delta$ and $\sigma$
should be measured at particular values in the future---to what extent is
one then still able to realize approximate textures in the neutrino Yukawa matrix?
Our analysis, thus, generalizes previous studies of the minimal 
seesaw model based on the assumption of exact texture zeros.
In particular, our study allows us to assess the theoretical uncertainty inherent
to the common texture ansatz.
One of our main results is that a normal light-neutrino mass
hierarchy is, in fact, still consistent with a two-zero Yukawa texture, provided that the two texture
zeros receive corrections at the level of $\mathcal{O}\left(\textrm{10}\,\%\right)$.
While our numerical results pertain to the minimal seesaw model only,
our general procedure appears to be applicable to other neutrino mass models as well.


\end{abstract}


\thispagestyle{empty}

\newpage



\section{Intro: Possible lessons from measuring leptonic
\texorpdfstring{\boldmath{$CP$}}{CP} violation?}
\label{sec:introduction}


After the celebrated discovery of neutrino masses and mixings during the last
two decades~\cite{Agashe:2014kda}, current and upcoming neutrino experiments
are now on the brink of taking the next big step: the observation
of $CP$ noninvariance in neutrino oscillations~\cite{Cabibbo:1977nk}.
Theory predicts that such leptonic $CP$ violation manifests itself in at least one
nonvanishing phase, $\delta$, in the lepton mixing matrix~\cite{Pontecorvo:1957qd};
and ongoing long-baseline experiments such as NO$\nu$A~\cite{Ayres:2004js}
and T2K~\cite{Abe:2011sj} prove to become increasingly sensitive to this observable.
Recently, the NO$\nu$A collaboration, e.g., reported their
newest constraints on the $CP$-violating phase $\delta$~\cite{Nova:2016aa},
which illustrate that NO$\nu$A is now capable of excluding
certain values of $\delta$ at the level of $3\,\sigma$.
Moreover, planned long-baseline experiments such as DUNE~\cite{Acciarri:2015uup}
will be able to actually measure $\delta$ over a large range of possible values.


From the viewpoint of theory, these recent developments as well as
experimental prospects directly lead to the question:
What can we possibly learn about the properties of the Standard Model (SM)
neutrinos from measuring a certain amount of $CP$ violation in
neutrino oscillations?
In this paper, we are going to attempt to answer this question
in the context of the minimal seesaw model, i.e.,
the type-I seesaw mechanism~\cite{seesaw} including only
two right-handed neutrinos.


The seesaw mechanism is arguably one of the most popular
explanations of the SM neutrino masses.
At the cost of a minimal extension of the SM particle
content, it manages to account for all observed low-energy
neutrino phenomena, while in addition offering the possibility
to explain the baryon asymmetry of the universe as a consequence of
baryogenesis via leptogenesis~\cite{Fukugita:1986hr}.
Traditionally, the seesaw mechanism supposes the existence
of three right-handed neutrinos, $N_I$ with $I=1,2,3$,
that transform as complete singlets under the SM gauge group.
In this standard scenario, all three SM neutrinos acquire
small nonvanishing masses, the differences between which are then
described by the observed solar and atmospheric mass-squared differences,
$\Delta m_{\rm sol}^2$ and $\Delta m_{\rm atm}^2$.
This realization of the seesaw mechanism, however, comes with
a large number of undetermined parameters at high energies,
which makes it difficult to derive distinct predictions
for the neutrino observables accessible at low energies.
A possible way out of this problem is to assume certain
(discrete) flavor symmetries that restrict the set of allowed
leptonic interactions in the Lagrangian, thereby reducing
the number of free parameters~\cite{Altarelli:2010gt}.
This approach is well motivated and has attracted (and continues
to attract) a great deal of attention in the literature.
On the other hand, the concept of flavor symmetries often times
requires a large number of additional scalar fields in order
to achieve certain Yukawa structures.
This increases the particle content and, hence, overall complexity
of the seesaw mechanism at high energies, which may make it even harder
to experimentally pin down its underlying dynamics at a fundamental level.


An intriguing and, in fact, quite orthogonal alternative to this approach
is to simply reduce the number of right-handed neutrinos
from three to two~\cite{King:1998jw,Frampton:2002qc}.
In this case, one has to deal with three complex Yukawa couplings
or, equivalently, six real parameters less.
As a consequence, one SM neutrino remains massless, while the masses of
the other two neutrinos are readily fixed by $\Delta m_{\rm sol}^2$ and
$\Delta m_{\rm atm}^2$.
This realization of the seesaw mechanism is, thus, significantly more
predictive than the standard scenario involving three right-handed neutrinos.
In addition, it nicely conforms with the philosophy of Occam's razor~\cite{Harigaya:2012bw},
which commands us to always give preference to explanations that rely
on the least number of assumptions. 
To explain the current low-energy neutrino
data, one actually does not \textit{need} to introduce three right-handed neutrinos.
Lacking knowledge of the absolute neutrino mass scale,
two right-handed neutrinos are absolutely sufficient.
Likewise, leptogenesis calls for two
right-handed neutrinos, but not necessarily more~\cite{Frampton:2002qc,Endoh:2002wm}.
Two (nondegenerate) right-handed neutrinos are sufficient to render
the decays of the heavy $N_I$ in the early universe $CP$-noninvariant,
as required by the Sakharov conditions for successful
baryogenesis~\cite{Sakharov:1967dj}.
A third right-handed neutrino is not essential.
In fact, as recently shown explicitly in Ref.~\cite{Bambhaniya:2016rbb}, 
the type-I seesaw mechanism with only two right-handed
neutrinos  succeeds in reproducing the observed baryon asymmetry of
the universe via \textit{resonant} leptogenesis~\cite{Flanz:1996fb}, i.e., for 
a nearly degenerate heavy-neutrino mass spectrum.%
\footnote{See also Ref.~\cite{Bjorkeroth:2016qsk} for a recent study of
leptogenesis in the supersymmetric type-I seesaw model.
In this model, two right-handed neutrino (super-) fields are sufficient
to realize inflation, leptogenesis, and SM neutrino masses.}
Taking into account arguments from electroweak naturalness, the authors
of Ref.~\cite{Bambhaniya:2016rbb} find that the two heavy neutrinos
in the minimal seesaw model should have a similar mass, not much larger
than $\mathcal{O}\left(10^7\right)\,\textrm{GeV}$.
In view of this result, we shall focus on more or less similar heavy-neutrino masses
in the following, neglecting the possibility of a strongly hierarchical spectrum.


In the case of two rather than three right-handed
neutrinos, the SM Lagrangian needs to be supplemented
with the following Yukawa and mass terms at high energies,
\begin{align}
\mathcal{L}_{\rm seesaw} = 
- y_{\alpha I}\, \ell_\alpha N_I H - \frac{1}{2}\, M_I N_I N_I + \textrm{h.c.}
\,, \quad \alpha = e,\mu,\tau
\,, \quad I = 1,2 \,.
\label{eq:Lseesaw}
\end{align}
Here, the first term accounts for the Yukawa
interactions of the right-handed neutrinos with the SM lepton and Higgs
doublets, $\ell_\alpha = \left(\nu_\alpha,e_\alpha\right)^T$ and $H = \left(H^+,H^0\right)^T$,
while the second term contains
Majorana masses, $M_I$, for the right-handed neutrinos, which, w.l.o.g., we
can take to be real and diagonal.
Meanwhile, the Yukawa couplings $y_{\alpha I}$
form an arbitrary complex $3\times2$ matrix.
The main goal of this paper will be to study the implications for
the Yukawa matrix $y_{\alpha I}$, provided that leptonic $CP$
violation should be conclusively observed in the future.


General seesaw models involving three right-handed neutrinos
predict the existence of, in total, three physical $CP$-violating phases
in the lepton mixing matrix:
one Dirac phase $\delta$, which is always present
and which may be regarded as the analog of the complex phase in the quark mixing
matrix~\cite{Kobayashi:1973fv}, as well as two additional phases, $\sigma$ and $\tau$,
owing to the Majorana nature of the SM neutrinos in the context of seesaw models.
In the case of only two right-handed neutrinos, one of these two Majorana phases
(say, $\tau$) can, however, be absorbed by a neutrino field rotation, so that only
two phases end up being physical: $\delta$ and $\sigma$.
Here, $\delta$ can be measured in neutrino oscillation experiments,
while $\sigma$ enters into the effective neutrino masses probed in searches
for neutrinoless double-beta decay (such as GERDA~\cite{Agostini:2013mzu}) or in experiments
aiming at a direct neutrino mass determination (such as KATRIN~\cite{Osipowicz:2001sq}).
In this paper, we are now going to study the possible hierarchy structures
and textures in the Yukawa matrix $y_{\alpha I}$ in dependence of
the $CP$-violating phases $\delta$ and $\sigma$.
To this end, we will adopt a two-step strategy:
First, we will introduce a convenient measure that characterizes
how close a given Yukawa matrix is to a certain flavor texture.
Then, we will perform a systematic parameter scan of the model in
Eq.~\eqref{eq:Lseesaw} and maximize, for fixed values of $\delta$
and $\sigma$, our \textit{hierarchy parameter} over all the unconstrained
parameters that are inaccessible to low-energy experiments.
This will allow us to assess how strong a hierarchy between different
Yukawa couplings one could possibly realize for any given values of
$\delta$ and $\sigma$. 
Our hope is that such a phenomenological link between experimentally accessible
observables on the one hand and fundamental theory parameters on the other hand
may help shed more light on the underlying dynamics
of the seesaw mechanism and on the UV origin of its flavor structure.
Our approach covers, in particular, all viable UV completions
of the minimal seesaw model from a bottom-up phenomenological perspective.


Our analysis ties in with studies of the minimal seesaw model that assume
exact texture zeros in the neutrino Yukawa 
matrix~\cite{Frampton:2002qc,Raidal:2002xf,King:2013iva,Zhang:2015tea}.
Such exact zeros may occur in consequence of certain flavor
symmetries in the fundamental high-energy theory of flavor.
Moreover, they comply with the spirit of Occam's razor and,
thirdly, they have proven to be a successful phenomenological
tool in the quark sector in the past.
As shown by Weinberg~\cite{Weinberg:1977hb}, imposing texture zeros in the mass matrices
for the first two quark generations allows one to derive a successful
prediction for the Cabibbo angle, $\theta_C = \left(m_d/m_s\right)^{1/2} \simeq 0.22$.
The main difference between our analysis and the ``texture ansatz''
is that we intend to tackle the flavor problem in the neutrino
sector from quite the opposite direction.
The philosophy behind neutrino mass models involving texture zeros
is to derive phenomenological predictions from a more or
less well motivated choice of Yukawa couplings at high energies.
This amounts to a classical top-down approach.
When confronted with the experimental data, such models typically 
lead to an \textit{all-or-nothing} situation:
Either the data turns out to be
consistent with the model's predictions or the model is ruled out.
We, by contrast, remain \textit{a priori} agnostic as to the structure
of the neutrino Yukawa matrix and let the data decide what hierarchies
or textures are experimentally preferred or ruled out.
In this sense, we are going to pursue a data-driven bottom-up approach,
which offers an interesting and conceptually new perspective on the interplay
between experimental and theoretical progress.
As we do not impose any conditions
on the neutrino Yukawa couplings from the outset, our approach
is considerably more model-independent than the conventional
theory-driven texture ansatz.
Or in other words, we no longer put all our eggs into one basket.
With the data being the driving force behind our analysis, any future
data update, if interpreted in the context of the minimal seesaw model,
is \textit{guaranteed} to provide us with new insights regarding
possible UV completions of the two-right-handed-neutrino model,
see also Fig.~\ref{fig:maps} in the Conclusions (Sec.~\ref{sec:conclusions}).


The rest of the paper is organized as follows.
In the next section, we shall first review the type-I seesaw mechanism
featuring two right-handed neutrinos.
In Sec.~\ref{sec:hierarchy}, we will then define our novel hierarchy parameter
that will allow us to quantify the quality of different hierarchies in
the neutrino Yukawa matrix.
Subsequently, we will present in Sec.~\ref{sec:scan} the results of our
scan of parameter space, i.e., the maximal hierarchy parameter
in dependence of $\delta$ and $\sigma$.
Here, our main conclusion will be that models with two right-handed
neutrinos and \textit{approximate} two-zero textures can still
accommodate both a normal (NH) and an inverse (IH) neutrino mass hierarchy.
This represents an important caveat to the conventional statement
that seesaw models with two right-handed neutrinos and two texture
zeros do not allow for a normally ordered light-neutrino mass spectrum.
Similarly, we will show how, in the case of an inverse
mass hierarchy, the viable set of exact textures can be generalized
to a much larger set of approximate textures.
Most of our observations are in accord with the naive
expectation that relaxing the assumption of an exact flavor texture
should enlarge the accessible part of parameter space.
But to us, it seems as if our approach presents the issue
of Yukawa hierarchies in minimal neutrino mass models from
a new angle and, thus, deserves a closer examination.
In a sense, it may be regarded as a supplement to all minimal
seesaw models that assume exact texture zeros in the neutrino Yukawa matrix. 
If one thinks of the predictions of these models as
``central values'' of a sort, our analysis yields the corresponding
``theoretical error bars'' to these predictions.
In Sec.~\ref{sec:stability}, we will briefly discuss the stability
of our numerical results under variations in the experimental
input data.
As we will see, our findings for $\delta$ and $\sigma$ exhibit an absolute
uncertainty owing to the experimental errors of at most $\mathcal{O}\left(1\,\%\right)\pi$.
In Sec.~\ref{sec:conclusions}, we will finally conclude
and give a brief outlook as to how our analysis could
be extended.
We will comment on the possibility of generalizing
our analysis to arbitrary textures as well as to the case of three right-handed neutrinos.


\section{Type-I seesaw mechanism with two right-handed neutrinos}
\label{sec:model}


Let us first recall how the seesaw Lagrangian in Eq.~\eqref{eq:Lseesaw} results
in small Majorana masses for the SM neutrinos.
In the course of electroweak symmetry breaking, the Higgs field $H^0$ acquires
a nonvanishing vacuum expectation value (VEV), $v_{\rm ew} \simeq 174\,\textrm{GeV}$.
This turns the Yukawa term in Eq.~\eqref{eq:Lseesaw} into a Dirac mass term
for the left-handed and right-handed neutrinos $\nu_\alpha$ and $N_I$,
\begin{align}
\mathcal{L}_{\rm seesaw} =
- \left(m_D\right)_{\alpha I}\, \nu_\alpha N_I - \frac{1}{2}\, M_I N_I N_I + \textrm{h.c.}
\,, \quad \left(m_D\right)_{\alpha I} = y_{\alpha I} \,v_{\rm ew} \,.
\end{align}
Provided there is a strong hierarchy between the Dirac and Majorana mass terms,
$M_I \gg \left(m_D\right)_{\alpha I}$, the right-handed neutrinos $N_I$
decouple at low energies, such that we can integrate them out.%
\footnote{From a technical point of view, this picture receives small
corrections once there is no longer a clear separation of scales, i.e.,
once the right-handed neutrino mass scale is sufficiently close to
the electroweak scale.
The active SM neutrinos then also mix with the heavy sterile neutrinos,
which renders the $3\times3$ lepton mixing matrix slightly non-unitary;
see, e.g., Ref.~\cite{Antusch:2014woa} and references therein.
These effects are, however, suppressed by the active-sterile mixing angle,
$\theta \sim m_D/M$, which is, in any case, bound to be tiny.
Given the experimental uncertainty of the currently known low-energy
neutrino observables, these corrections are, hence, always negligible
for our purposes.}
This results in the famous type-I seesaw formula for the SM neutrino masses,
\begin{align}
\mathcal{L}_{\rm seesaw} =
- \left(m_\nu\right)_{\alpha \beta}\, \nu_\alpha \nu_\beta + \textrm{h.c.}
\,, \quad \left(m_\nu\right)_{\alpha \beta} =
- v_{\rm ew}^2\, y_{\alpha I}\, M_I^{-1}\, y_{\beta I} \,.
\label{eq:mnu}
\end{align}
Here, the SM neutrino mass matrix, $\left(m_\nu\right)_{\alpha \beta}$,
is a complex symmetric $3\times3$ matrix, which needs to be diagonalized
by a unitary Takagi factorization.
In matrix notation, this means
\begin{align}
D_\nu = U^T\,m_\nu\,U \,, \quad D_\nu = \textrm{diag}\left\{m_1,m_2,m_3\right\} \,.
\label{eq:Dnu}
\end{align}
For a normally ordered light-neutrino mass spectrum, we have
$m_1 < m_2 < m_3$, whereas an inversely ordered
light-neutrino  mass spectrum is characterized
by $m_3 < m_1 < m_2$.
In the presence of only two right-handed neutrinos,
$\left(m_\nu\right)_{\alpha \beta}$ in Eq.~\eqref{eq:mnu} is a rank-2 matrix.
The lightest neutrino mass eigenvalue therefore
vanishes in our model for both NH and IH, $\min\left\{m_i\right\} = 0$.


We shall work in a basis in which the charged-lepton mass matrix
is diagonal.
Then we can identify the unitary matrix $U$ in Eq.~\eqref{eq:Dnu}
as the lepton mixing or Pontecorvo-Maki-Nakagawa-Sakata (PMNS) matrix,
which relates the flavor eigenstates $\nu_\alpha$
to the mass eigenstates $\nu_i$,
\begin{align}
\nu_\alpha = U_{\alpha i}\, \nu_i \,, \quad
\nu_i = U_{\alpha i}^* \, \nu_\alpha \,, \quad
\alpha = e,\mu,\tau \,, \quad i = 1,2,3 \,.
\end{align}
In the case of one massless SM neutrino,
the PMNS matrix encompasses five (instead of the usual six)
physical degrees of freedom (DOFs):
three mixing angles, $\theta_{12}$, $\theta_{13}$, and $\theta_{23}$,
which take values in the interval $[0,\pi/2]$,
as well as two $CP$-violating phases, $\delta \in \left[0,2\pi\right)$
and $\sigma \in \left[0,\pi\right)$,
\begin{align}
U = \begin{pmatrix}
c_{12}\,c_{13} &
s_{12}\,c_{13} &
s_{13}\,e^{-i\delta} \\
-s_{12}\,c_{23} - c_{12}\,s_{13}\,s_{23}\,e^{i\delta} &
\phantom{-}c_{12}\,c_{23} - s_{12}\,s_{13}\,s_{23}\,e^{i\delta} &
s_{23}\,c_{13} \\
\phantom{-} s_{12}\,s_{23} - c_{12}\,s_{13}\,c_{23}\,e^{i\delta} & 
-c_{12}\,s_{23} - s_{12}\,s_{13}\,c_{23}\,e^{i\delta} & 
c_{23}\,c_{13}
\end{pmatrix}
\begin{pmatrix}
1 & 0 & 0 \\
0 & e^{i\sigma} & 0 \\
0 & 0 & 1
\end{pmatrix} \,.
\end{align}
Here, $c_{ij}$ and $s_{ij}$ are a shorthand notation for $\cos \theta_{ij}$
and $\sin \theta_{ij}$, respectively.
In total, we, thus, arrive at the conclusion that the SM neutrino sector
features seven low-energy observables: the five observables encoded
in the PMNS matrix as well as two nonzero mass eigenvalues.
This needs to be contrasted with the available parameters at high energies.
The neutrino Yukawa matrix $y_{\alpha I}$ contains twelve real parameters---six
absolute values and six phases---out of which three phases are unphysical as they
can be absorbed by charged-lepton rotations.
Meanwhile, the right-handed neutrino masses $M_{1,2}$ constitute two further
high-energy parameters.
Changes in these masses can, however, be compensated for by a rescaling of
the neutrino Yukawa couplings,
\begin{align}
M_I \rightarrow M_I' \,, \quad 
y_{\alpha I} \rightarrow \left(\frac{M_I'}{M_I}\right)^{1/2} y_{\alpha I} \,.
\label{eq:repara}
\end{align}
This leaves us with, in total, $12 - 3 + 2 - 2 = 9$ independent parameters
at high energies, which is sufficient to account for the seven
low-energy observables contained in $D_\nu$ and $U$.


The functional relation between the input parameters at high energies
and the observables at low energies becomes most transparent
in the Casas-Ibarra parametrization of the neutrino Yukawa matrix~\cite{Casas:2001sr}.
To re-derive this parametrization making use of our conventions and notations,
we simply have to combine the seesaw formula in Eq.~\eqref{eq:mnu} with
the relation in Eq.~\eqref{eq:Dnu},
\begin{align}
D_\nu = - v_{\rm ew}^2\, U^T\,y\,D_N^{-1}\,y^T\,U \,, \quad
D_N = \textrm{diag}\left\{M_1,M_2\right\} \,.
\end{align}
Solving this matrix equation for the neutrino Yukawa matrix, one finds
\begin{align}
y = \frac{i}{v_{\rm ew}}\, U^* D_\nu^{1/2}\,R\,D_N^{1/2} \,,
\label{eq:CIpara}
\end{align}
where $R$ is a complex $3\times2$ rotation matrix that satisfies $R^TR = \mathbb{1}_{2\times2}$,
but \textit{not} $RR^T = \mathbb{1}_{3\times3}$,
\begin{align}
\textrm{NH:} \quad m_1 = 0 \,, \quad
R = \begin{pmatrix}
0 & 0 \\
\cos z & -\sin z \\
\zeta\sin z &  \zeta\cos z
\end{pmatrix} \,, \quad
\textrm{IH:} \quad m_3 = 0 \,, \quad
R = \begin{pmatrix}
\cos z & -\sin z \\
\zeta\sin z & \zeta\cos z \\
0 & 0
\end{pmatrix} \,.
\end{align}
Here, the rotation angle $z$ is an arbitrary complex number, which represents
the two real excess DOFs in the high-energy Yukawa matrix $y_{\alpha I}$
compared to the low-energy mass matrix $\left(m_\nu\right)_{\alpha\beta}$.
The parameter $\zeta=\pm 1$ distinguishes between a ``positive'' and a
``negative'' branch of possible rotation matrices $R$.
The resulting Yukawa matrices on both branches can, however, be mapped onto each
other by a combination of sign changes, column exchange, etc.;
see Appendix~B of Ref.~\cite{Bjorkeroth:2016qsk} for details.
For our purposes, we are, thus, free to
either pick $\zeta = + 1$ or $\zeta = -1$.
In the following, we will work, w.l.o.g., with the Yukawa
matrices on the positive branch, $\zeta = + 1$.
The Casas-Ibarra parametrization also exhibits
the re-parametrization freedom displayed in Eq.~\eqref{eq:repara}.
To see this explicitly, it turns out convenient to introduce
the following quantities,
\begin{align}
\label{eq:kappaV}
\kappa_{\alpha I} = -i\,y_{\alpha I}\,\sqrt{\frac{v_{\rm ew}}{M_I}} \,, \quad
V_{\alpha i} = U_{\alpha i}^*\,\sqrt{\frac{m_i}{v_{\rm ew}}} \,,
\end{align}
which allow us to write down a dimensionless version of the Casas-Ibarra
parametrization,
\begin{align}
\textrm{NH:} \quad
\begin{pmatrix}
\kappa_{\alpha 1} \\
\kappa_{\alpha 2}
\end{pmatrix} =
\begin{pmatrix}
 \cos z & \sin z \\
-\sin z & \cos z 
\end{pmatrix}
\begin{pmatrix}
V_{\alpha2} \\
V_{\alpha3}
\end{pmatrix} \,, \quad
\textrm{IH:} \quad
\begin{pmatrix}
\kappa_{\alpha 1} \\
\kappa_{\alpha 2}
\end{pmatrix} =
\begin{pmatrix}
 \cos z & \sin z \\
-\sin z & \cos z 
\end{pmatrix}
\begin{pmatrix}
V_{\alpha1} \\
V_{\alpha2}
\end{pmatrix} \,.
\label{eq:CIpara0}
\end{align}
The expressions on the right-hand sides of these two equations now no longer
depend on the heavy-neutrino masses $M_{1,2}$.
Instead, these two masses are absorbed in the rescaled Yukawa
couplings $\kappa_{\alpha I} \propto y_{\alpha I} M_I^{-1/2}$, which
are invariant under the scaling transformation in Eq.~\eqref{eq:repara}.
Note also that this version of the Casas-Ibarra parametrization
nicely separates the unknown high-energy input parameters
(on the left-hand side) from the low-energy observables
that can be measured in experiments (on the right-hand side).
As evident from Eq.~\eqref{eq:CIpara0}, both sets of quantities are related 
by a rotation about a complex angle $z$.
This angle is physically meaningful at high energies,
but appears as an ``unphysical'' auxiliary parameter at low energies.


An important implication of Eq.~\eqref{eq:CIpara0} is that we can
always find values of the complex parameter $z$, such that at least
one Yukawa coupling vanishes exactly.
In fact, all we have to do is to choose $z$ such that $\tan z$
equals the ratio of two entries in the rescaled PMNS matrix $V_{\alpha i}$,
\begin{align}
\label{eq:T1zero}
\textrm{NH:} \quad
\kappa_{\alpha 1} & = 0 \quad\Rightarrow & \mkern-18mu\tan z & = -\frac{V_{\alpha 2}}{V_{\alpha 3}} \,, &
\textrm{IH:} \quad
\kappa_{\alpha 1} & = 0 \quad\Rightarrow & \mkern-18mu\tan z & = -\frac{V_{\alpha 1}}{V_{\alpha 2}} \,,\\ \nonumber
\kappa_{\alpha 2} & = 0 \quad\Rightarrow & \mkern-18mu\tan z & = +\frac{V_{\alpha 3}}{V_{\alpha 2}} \,, &
\kappa_{\alpha 2} & = 0 \quad\Rightarrow & \mkern-18mu\tan z & = +\frac{V_{\alpha 2}}{V_{\alpha 1}} \,.
\end{align}
Requiring a one-zero texture in the neutrino Yukawa matrix, thus, eliminates
the parameter $z$ and leaves us with an equal number of input parameters and
observables.
Such one-zero texture models have been extensively discussed in the
literature~\cite{King:2013iva}, more recently in particular in the context
of \textit{constrained sequential dominance}~\cite{King:2005bj};
see, e.g., Ref.~\cite{Bjorkeroth:2015tsa} and references therein.


Realizing two zeros in the neutrino Yukawa matrix is already more challenging,
as it requires a conspiracy among four entries in the rescaled PMNS matrix $V_{\alpha i}$.
Schematically, we have
\begin{align}
\left(\kappa_{\alpha I},\kappa_{\beta J}\right) = \left(0,0\right)
\,, \quad \left(\alpha,I\right) \neq \left(\beta,J\right)
\quad\Rightarrow\quad
\frac{V_1}{V_2} = \pm
\frac{V_3}{V_4} \,,
\label{eq:T2zero}
\end{align}
with the exact matrix entries $V_{1,2,3,4}$
following from Eq.~\eqref{eq:T1zero}, respectively.
In fact, such scenarios correspond to the most minimal realizations of the
seesaw mechanism~\cite{Frampton:2002qc,Harigaya:2012bw}.
It is easy to see that, in the case of only two right-handed neutrinos,
three or more texture zeros in the neutrino Yukawa matrix
lead to at least two vanishing neutrino mixing angles.
Therefore, given the fact that all three mixing angles, including the reactor mixing
angle $\theta_{13}$, have been measured to be nonzero in experiments in recent
years, models with more than two texture zeros clearly conflict with the data.
Meanwhile, also models with two texture zeros are highly constrained.
To illustrate the range of viable two-zero textures, let us adopt
the notation of Ref.~\cite{Zhang:2015tea},
which categorizes all possible two-zero textures in the matrix $y_{\alpha I}$
into three groups, $A_i$, $B_i$, and $C_i$,


\begin{align}
\label{eq:ABC}
& A_1 \,:\:
\begin{pmatrix}
0 & 0           \\
\times & \times \\
\times & \times 
\end{pmatrix}\,, \quad 
A_2 \,:\:
\begin{pmatrix}
\times & \times \\
0 & 0           \\
\times & \times 
\end{pmatrix}\,, \quad 
A_3 \,:\:
\begin{pmatrix}
\times & \times \\
\times & \times \\
0 & 0     
\end{pmatrix}\,, \\ \nonumber
& B_1 \,:\:
\begin{pmatrix}
0 & \times      \\
\times & 0      \\
\times & \times 
\end{pmatrix}\,, \quad 
B_2 \,:\:
\begin{pmatrix}
0 & \times      \\
\times & \times \\
\times & 0
\end{pmatrix}\,, \quad 
B_3 \,:\:
\begin{pmatrix}
\times & \times \\
0 & \times      \\
\times & 0     
\end{pmatrix}\,, \\ \nonumber
& B_4 \,:\:
\begin{pmatrix}
\times & 0      \\
0 & \times      \\
\times & \times 
\end{pmatrix}\,, \quad 
B_5 \,:\:
\begin{pmatrix}
\times & 0      \\
\times & \times \\
0 & \times 
\end{pmatrix}\,, \quad 
B_6 \,:\:
\begin{pmatrix}
\times & \times \\
\times & 0      \\
0 & \times     
\end{pmatrix}\,, \\ \nonumber
& C_1 \,:\:
\begin{pmatrix}
0 & \times      \\
0 & \times      \\
\times & \times 
\end{pmatrix}\,, \quad 
C_2 \,:\:
\begin{pmatrix}
0 & \times      \\
\times & \times \\
0 & \times 
\end{pmatrix}\,, \quad 
C_3 \,:\:
\begin{pmatrix}
\times & \times \\
0 & \times      \\
0 & \times
\end{pmatrix}\,, \\ \nonumber
& C_4 \,:\:
\begin{pmatrix}
\times & 0      \\
\times & 0      \\
\times & \times 
\end{pmatrix}\,, \quad 
C_5 \,:\:
\begin{pmatrix}
\times & 0      \\
\times & \times \\
\times & 0
\end{pmatrix}\,, \quad 
C_6 \,:\:
\begin{pmatrix}
\times & \times \\
\times & 0      \\
\times & 0     
\end{pmatrix}\,.
\end{align}
\smallskip


As shown in Refs.~\cite{Harigaya:2012bw,Zhang:2015tea}, only four out of these
15 textures turn out to be compatible with the current low-energy neutrino data:
$B_1$, $B_2$, $B_4$, and $B_5$---and that even only in the case of an
inverse mass hierarchy.
This amounts to the statement that the type-I seesaw model with
two right-handed neutrinos and an \textit{exact} two-zero texture
does not allow for a normally ordered light-neutrino mass spectrum.
Of course, this conclusion no longer holds true once one gives up on
the texture zeros in Eq.~\eqref{eq:ABC} and replaces all zeros by nonzero entries.
One of our main goals in this paper will be to quantify this
statement and to assess how large these corrections to an exact two-zero
texture need to be so that a normal mass hierarchy comes
into reach again.
Moreover, it is important to note that the pairs of textures
$B_1$ and $B_4$ as well as $B_2$ and $B_5$ only differ by
an exchange of the two columns in $y_{\alpha I}$. 
Physically, this simply corresponds to
the two neutrinos $N_1$ and $N_2$ interchanging their coupling
strengths to the SM lepton-Higgs fields.
Such a change in the Yukawa term in Eq.~\eqref{eq:Lseesaw} can,
however, always be compensated for by a corresponding change in
the $N_{1,2}$ Majorana masses, $M_1 \leftrightarrow M_2$.
The textures $B_{1,4}$ and $B_{2,5}$ are, thus, bound
to lead to the same low-energy predictions, respectively.
This is best seen in the predictions for the $CP$-violating phases
$\delta$ and $\sigma$. 
Owing to the restrictive condition in
Eq.~\eqref{eq:T2zero}, these two observables are completely determined
by the neutrino masses and mixings angles.
To leading order in the small mixing angle $s_{13}$, one finds~\cite{Zhang:2015tea}
\begin{align}
\label{eq:dspred}
\cos \delta \simeq
\begin{cases}
+ \frac{\sin 2\theta_{12}\left(1-m_1^2/m_2^2\right)}{4\tan\theta_{23}\sin\theta_{13}}
- \frac{\tan\theta_{23}\sin\theta_{13}}{\tan2\theta_{12}}  \\          
- \frac{\sin 2\theta_{12}\left(1-m_1^2/m_2^2\right)}{4\cot\theta_{23}\sin\theta_{13}}
+ \frac{\cot\theta_{23}\sin\theta_{13}}{\tan2\theta_{12}} 
\end{cases} \hspace{-8pt} \,, \quad
\cos 2\sigma \simeq
\begin{cases}
1 - \frac{\tan^2\theta_{23}\sin^2\theta_{13}}{2\sin^2\theta_{12}\cos^2\theta_{12}} &
\quad \left(B_{1,4}\right) \\          
1 - \frac{\cot^2\theta_{23}\sin^2\theta_{13}}{2\sin^2\theta_{12}\cos^2\theta_{12}} &
\quad \left(B_{2,5}\right)
\end{cases} \,.
\end{align}


\begin{table}
\begin{center}
\begin{tabular}{|c||cccc|}\hline
Observable & Units & Hierarchy & Best-fit value & $3\,\sigma$ confidence interval \\\hline\hline
$\delta m^2$ & $\left[10^{-5}\,\textrm{eV}^2\right]$ & both & $+7.37$ & $\left[+6.93,+7.97\right]$ \\\hline
\multirow{2}{*}{$\Delta m^2$} &
\multirow{2}{*}{$\left[10^{-3}\,\textrm{eV}^2\right]$}
& NH & $+2.50$ & $\left[+2.37,+2.63\right]$ \\
& & IH & $-2.46$ & $\left[-2.33, -2.60\right]$ \\\hline
$\sin^2\theta_{12}$ & $\left[10^{-1}\right]$ & both & $+2.97$ & $\left[+2.50,+3.54\right]$ \\\hline
\multirow{2}{*}{$\sin^2\theta_{13}$} &
\multirow{2}{*}{$\left[10^{-2}\right]$}
& NH & $+2.14$ & $\left[+1.85,+2.46\right]$ \\
& & IH & $+2.18$ & $\left[+1.86,+2.48\right]$ \\\hline
\multirow{2}{*}{$\sin^2\theta_{23}$} &
\multirow{2}{*}{$\left[10^{-1}\right]$}
& NH & $+4.37$ & $\left[+3.79,+6.16\right]$ \\
& & IH & $+5.69$ & $\left[+3.83,+6.37\right]$ \\\hline
\multirow{2}{*}{$\delta$} &
\multirow{2}{*}{$\left[\pi\right]$}
& NH & $+1.35$ & $\left[+0.00,+2.00\right]$ \\
& & IH & $+1.32$ & $\left[+0.00,+2.00\right]$  \\\hline
\end{tabular}
\caption{Best-fit values and $3\,\sigma$ confidence intervals for the five low-energy
observations that are currently accessible in experiments
($\delta m^2$, $\Delta m^2$, $\sin^2\theta_{12}$,
$\sin^2\theta_{13}$, and $\sin^2\theta_{23}$) as well as for the $CP$-violating phase
$\delta$~\cite{Capozzi:2016rtj}.
Note that, at the $3\,\sigma$ level, $\delta$ can still freely vary over the entire range
of possible values, $\delta \in \left[0,2\pi\right)$.}
\label{tab:data}
\end{center}
\end{table}


These predictions can be evaluated by making use of the experimental results
for the observed masses and mixings.
Among the various global-fit results in the literature~\cite{Forero:2014bxa,Capozzi:2016rtj},
we shall use the most recent analysis~\cite{Capozzi:2016rtj} in this paper,
for definiteness.
Ref.~\cite{Capozzi:2016rtj} uses the global neutrino data from several
neutrino observatories, accelerators and reactors to constrain
six observables: $\delta m^2$, $\Delta m^2$, $\sin^2\theta_{12}$,
$\sin^2\theta_{13}$, $\sin^2\theta_{23}$, and $\delta$.
Here, the two mass-squared differences $\delta m^2$ and $\Delta m^2$
correspond to more accurately defined versions of $\Delta m_{\rm sol}^2$
and $\Delta m_{\rm atm}^2$, respectively,
\begin{align}
\delta m^2 = m_2^2 - m_1^2 \,, \quad
\Delta m^2 = m_3^2 - \frac{1}{2}\left(m_2^2 + m_1^2\right) \,.
\end{align}
Given that, in our scenario, $m_1 = 0$ in the NH case and $m_3 = 0$ in the IH case,
the three light-neutrino masses $m_{1,2,3}$ can then be uniquely expressed
in terms of $\delta m^2$ and $\Delta m^2$ as follows,
\begin{align}
\textrm{NH:} \quad
m_1 & = 0 \,, &
\textrm{IH:} \quad
m_1 & = \left(-\Delta m^2 - \frac{1}{2}\delta m^2\right)^{1/2} \,,\\ \nonumber
m_2 & = \left(\delta m^2\right)^{1/2} \,, &
m_2 & = \left(-\Delta m^2 + \frac{1}{2}\delta m^2\right)^{1/2} \,,\\ \nonumber
m_3 & = \left(\Delta m^2 + \frac{1}{2}\delta m^2\right)^{1/2} \,, &
m_3 & = 0 \,.
\end{align}
Here, note that the sign of $\Delta m^2$ is not fixed:
$\Delta m^2 > 0$ for NH and $\Delta m^2 < 0$ for IH.
The best-fit values
as well as $3\,\sigma$ confidence intervals for
all six observables are summarized in Tab.~\ref{tab:data}.
With the aid of the best-fit values for $\delta m^2$, $\Delta m^2$, $\sin^2\theta_{12}$,
$\sin^2\theta_{13}$, and $\sin^2\theta_{23}$, we then
find the following predictions for the phases $\delta$ and $\sigma$ in
models with an exact two-zero texture,
\begin{align}
\label{eq:dsB1245}
\frac{\left(\delta,\sigma\right)}{\pi} \simeq
\begin{cases}
\left(0.51,0.94\right) \quad\textrm{or}\quad \left(1.49,0.06\right)
& \qquad \left(B_{1,4}\right) \\          
\left(0.50,0.04\right) \quad\textrm{or}\quad \left(1.50,0.96\right)
& \qquad \left(B_{2,5}\right)
\end{cases} \,.
\end{align}
A combined determination of $\delta$ and $\sigma$ in experiment (or a sole determination
of $\delta$ with a precision at the percent level) would,
thus, allow to distinguish between the two possible sets of Yukawa textures.
Moreover, it is interesting to observe that both models predict
a $CP$-violating phase $\delta$ close to its maximal value,
$\delta \simeq \pm \pi/2 \:\textrm{mod}\: 2\pi$.
Remarkably enough, this is in agreement with the (arguably weak)
trend in the latest experimental constraints on $\delta$
(see also Ref.~\cite{Nova:2016aa}).


\section{Assessing hierarchies in the neutrino Yukawa matrix}
\label{sec:hierarchy}


A clear advantage of models with an exact two-zero texture is their testability.
If the Yukawa matrix $y_{\alpha I}$ should, indeed, contain two exact zeros,
experiments should measure $\delta$ and $\sigma$ at values not much different
from those in Eq.~\eqref{eq:dsB1245}.
However, we are left wondering: What if this should turn out not to be the case?
After all, the predictions in Eq.~\eqref{eq:dsB1245} offer us an
\textit{all-or-nothing} perspective:
Either experiments determine the phases $\delta$ and $\sigma$ at the predicted
values or the minimal seesaw model with two exact texture zeros
is immediately ruled out! 
So, if future data should indicate some deviations from the exact
predictions in Eq.~\eqref{eq:dsB1245}, should we then abandon
the idea of two-zero textures altogether?
At this point, we caution that one should not jump to
conclusions too quickly.
The predictions in Eq.~\eqref{eq:dsB1245} may have experimental
error bars (see our analysis in Sec.~\ref{sec:stability} and, in particular,
Fig.~\ref{fig:expuncert}); but from a theoretical perspective,
they are extremely stiff.
Not only does this raise the stakes in the \textit{all-or-nothing} game,
it may also appear unnatural for a couple of physical reasons.
In reality, the picture of exact texture zeros may very well receive corrections
that perturb the naive leading-order neutrino Yukawa matrix.
Two exact texture zeros could easily be nothing but a first-order
approximation and the matrix $y_{\alpha I}$ may, in fact, rather exhibit a hierarchy
structure in the sense that two of its entries are parametrically smaller
than the other four.
In the high-energy theory of flavor, this may, e.g., be achieved via some sort of
explicit flavor symmetry breaking.
Alternatively, the texture in the Yukawa matrix may only be exact at tree level
and in the absence of gravity, such that radiative and/or gravitational
effects might be responsible for small perturbations.%
\footnote{At this point, it is worthwhile recalling that quantum
gravity is expected (on very general grounds) to explicitly break any
global symmetry~\cite{Banks:2010zn}.
Therefore, if it should be a \textit{global} flavor symmetry that is
responsible for the (approximate) two-zero texture, small perturbations
of the no-gravity approximation are almost unavoidable.\smallskip}
In this paper, we will, however, not delve any further into such speculations.
Instead, we are going to study the possibility of \textit{approximate}
Yukawa textures from a bottom-up phenomenological point of view.
We would like to address the question as to how strong a hierarchy
could still be realized in $y_{\alpha I}$, given that 
$\delta$ and $\sigma$ should be measured at values different from those
in Eq.~\eqref{eq:dsB1245}.
Among other things, this will provide us with theoretical error bars
on the predictions in Eq.~\eqref{eq:dsB1245}, which we deem useful and relevant
not only from a systematic, but also from a physical perspective.


To achieve our goal, we need to compare with each other all possible neutrino Yukawa matrices
that lead to the same predictions for $\delta$ and $\sigma$ and find, for each pair
of $\left(\delta,\sigma\right)$ values, the most
``extreme'' matrix that exhibits the strongest hierarchy.
Fortunately, such an endeavor is greatly facilitated by the Casas-Ibarra
parametrization in Eq.~\eqref{eq:CIpara0}.
For given values of $\delta$ and $\sigma$, we simply need to evaluate
the $V_{\alpha i}$ matrix elements on the right-hand sides of the two
relations in Eq.~\eqref{eq:CIpara0}.
The respective sets of Yukawa couplings that we are interested in
then readily follow from varying the complex parameter $z$ over its domain.%
\footnote{In this sense, each viable UV completion of the minimal seesaw model
can be associated with a certain $z$ value.
Scanning the entire complex $z$ plane, we therefore make sure to cover \textit{all}
viable UV completions of the minimal seesaw model, irrespectively of whether
these have been explicitly worked out in the literature or not.}
In other words, by following this procedure, we find a different Yukawa
matrix at each point in the complex $z$ plane; but all of the matrices
constructed in this way lead to the same predictions for $\delta$ and $\sigma$.
For each matrix, we can then examine the relation between the different
Yukawa couplings.
To do so, let us sort the six elements of a given matrix according
to the size of their absolute values,
\begin{align}
\kappa = \begin{pmatrix}
\kappa_{e1}    & \kappa_{e2}    \\
\kappa_{\mu1}  & \kappa_{\mu2}  \\
\kappa_{\tau1} & \kappa_{\tau2} 
\end{pmatrix} \quad\rightarrow\quad
\hat{\kappa} = \left(\hat{\kappa}_1,\hat{\kappa}_2,\hat{\kappa}_3,
\hat{\kappa}_4,\hat{\kappa}_5,\hat{\kappa}_6\right) \,, \quad
\left|\hat{\kappa}_\mu\right| \leq \left|\hat{\kappa}_{\mu+1}\right| \,, 
\quad \mu = 1, \cdots,5 \,.
\end{align}
Texture zeros in the Yukawa matrix then manifest itself in the first
elements of the sorted list $\hat{\kappa}$ being zero,
independently of where in the Yukawa matrix the texture zeros appear,
\begin{align}
\textrm{One texture zero:} & \quad \hat{\kappa}_1 = 0 \,, \quad \hat{\kappa}_2 > 0
\,, \\ \nonumber
\textrm{Two texture zeros:} & \quad \hat{\kappa}_2 = 0 \,, \quad \hat{\kappa}_3 > 0 \,,
\end{align}
and so on.
In the following, we shall particularly focus on approximate two-zero
textures in the neutrino Yukawa matrix.
As discussed in the previous section, one-zero textures can be trivially
realized for arbitrary values of $\delta$ and $\sigma$, while three-zero
textures lead to unacceptable predictions for the neutrino mixing angles.
An \textit{approximate} two-zero textures is then characterized by
the fact that $\left|\hat{\kappa}_{1,2}\right|$ are much smaller than
$\left|\hat{\kappa}_{3,4,5,6}\right|$.
This motivates us to define the following \textit{hierarchy parameter} $R_{23}$
as a measure for the gap in between these two subsets of Yukawa couplings,
\begin{align}
\label{eq:R23}
R_{23}\left(\delta,\sigma;z\right) =
\frac{\left|\hat{\kappa}_2\right|}{\left|\hat{\kappa}_3\right|} \,.
\end{align}


Let us study the properties of this parameter in a bit more detail.
First of all, we note that, while $R_{23}$ is defined in terms of rescaled
Yukawa couplings $\kappa_{\alpha I}$, it can also be rewritten in terms
of actual Yukawa couplings $y_{\alpha I}$.
According to Eq.~\eqref{eq:kappaV}, we can define
$\hat{y}_{2,3} = i \big(\hat{M}_{2,3} / v_{\rm ew}\big)^{1/2}\hat{\kappa}_{2,3}$,
where $\hat{M}_{2,3}$ either needs to be identified as $M_1$ or $M_2$.
$R_{23}$ then takes the following form,
\begin{align}
R_{23} = \left(\frac{\hat{M}_3}{\hat{M}_2}\right)^{1/2}
\frac{\left|\hat{y}_2\right|}{\left|\hat{y}_3\right|} \,.
\label{eq:R23y}
\end{align}
That is, $R_{23}$ corresponds to a ratio of actual Yukawa couplings,
$\left|\hat{y}_2\right| / \left|\hat{y}_3\right|$,
rescaled by the square root of a ratio of heavy-neutrino mass eigenvalues,
$\big|\hat{M}_3\big| / \big|\hat{M}_2\big|$.
In the following, we shall assume that the two heavy-neutrino
masses $M_1$ and $M_2$ do not exhibit a strong hierarchy
(say, $1/3 \lesssim M_1 / M_2 \lesssim 3$).
This is inspired by the observation that successful leptogenesis
in the minimal seesaw model relies on a nearly degenerate
heavy-neutrino mass spectrum, $M_1 \simeq M_2$~\cite{Bambhaniya:2016rbb}.
For $1/3 \lesssim M_1 / M_2 \lesssim 3$,
the mass ratio in Eq.~\eqref{eq:R23y} then becomes more or less negligible,
\begin{align}
\label{eq:massassump}
\frac{M_1}{M_2} \sim \mathcal{O}(1) \quad\Rightarrow\quad
R_{23} = \frac{\left|\hat{\kappa}_2\right|}{\left|\hat{\kappa}_3\right|} \simeq
\frac{\left|\hat{y}_2\right|}{\left|\hat{y}_3\right|} \,.
\end{align}
Therefore, once $M_1$ and $M_2$ are roughly of the
same order of magnitude, $R_{23}$ truly corresponds to a ratio of elements
in the actual (and not only rescaled) neutrino Yukawa matrix $y_{\alpha I}$.


On the other hand, we emphasize that the parameter $R_{23}$ is defined
in Eq.~\eqref{eq:R23} in terms of the rescaled Yukawa couplings $\hat{\kappa}_2$
and $\hat{\kappa}_3$ \textit{on purpose}.
Thanks to this definition, all of our numerical results in the following
will be independent of the concrete values of $M_{1,2}$, which renders
our approach more model-independent.
For a specific choice of heavy-neutrino mass eigenvalues,
our results can then be trivially converted into statements about
the ratio of actual Yukawa couplings by making use of Eq.~\eqref{eq:R23y}.
Here, it should be noted, however, that our conclusions regarding exact texture zeros
will not be affected by the explicit values of $M_{1,2}$. 
Exact zeros in the rescaled matrix $\kappa_{\alpha I}$ translate into exact zeros
in the matrix $y_{\alpha I}$, see Eq.~\eqref{eq:kappaV}.
For all other textures, the rescaling in Eq.~\eqref{eq:R23y} leads to a universal
suppression or enhancement.

\begin{figure}
\begin{center}
\includegraphics[width=0.475\textwidth]{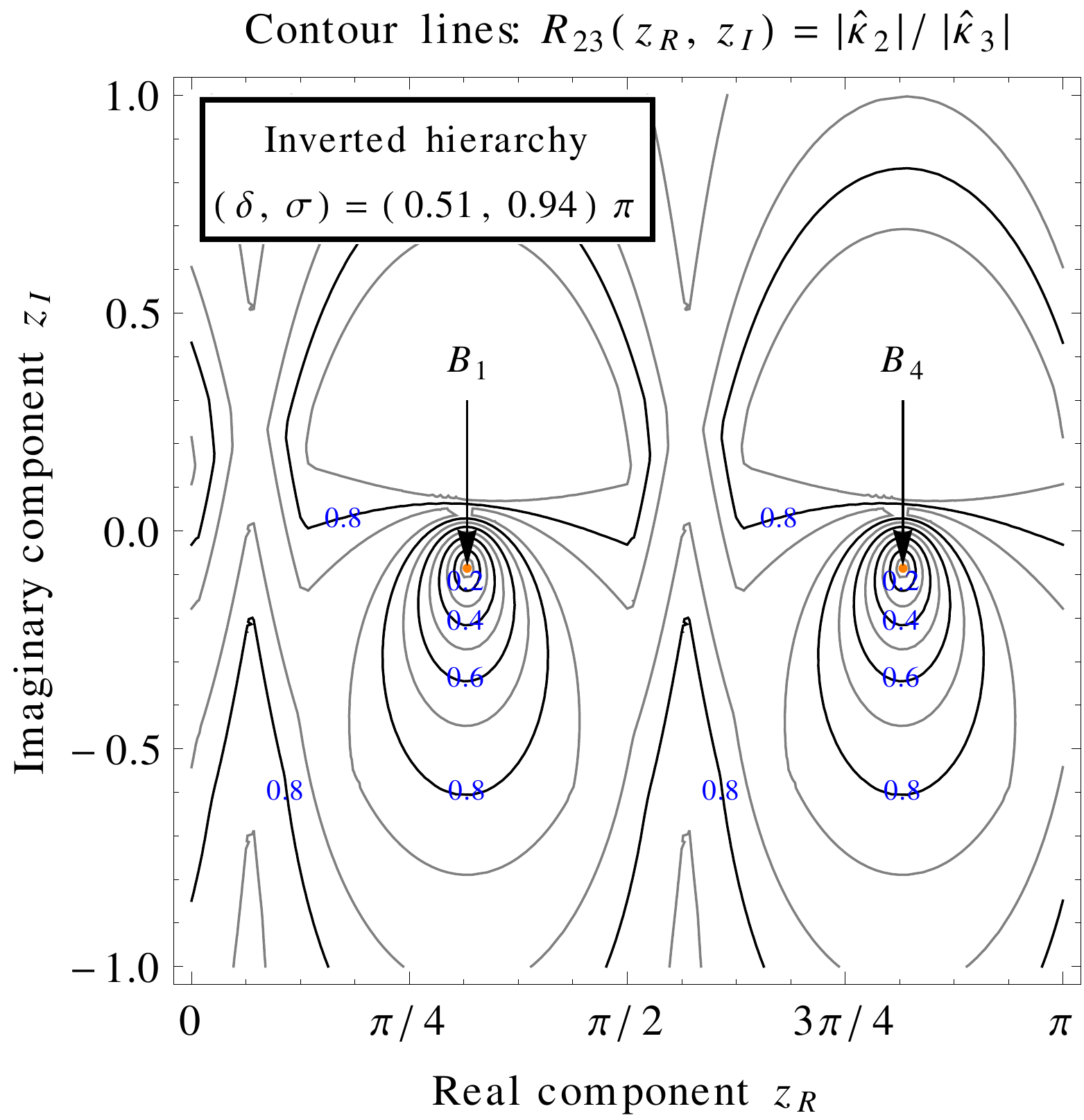}\hfill
\includegraphics[width=0.475\textwidth]{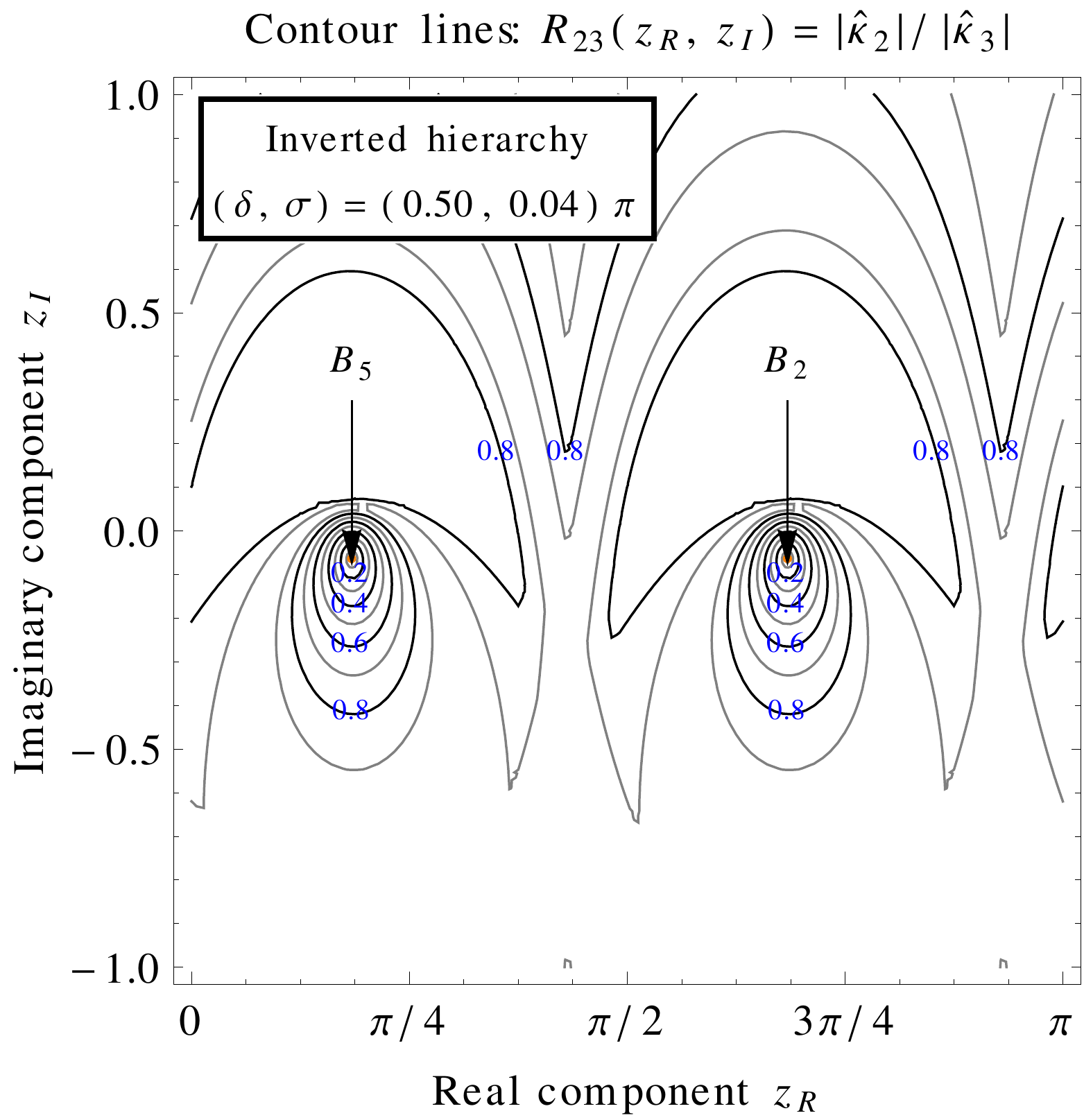}
\caption{Hierarchy parameter $R_{23}$ as a function of the complex parameter
$z$ after setting the phases $\delta$ and $\sigma$ to their
predicted values in the $B_{1,4}$ \textbf{(left panel)}
and $B_{2,5}$ \textbf{(right panel)} scenarios, respectively
(see Eq.~\eqref{eq:dsB1245}).
Notice that the values of $z$ leading to exact two-zero textures
correspond to zeros of $R_{23}$ in the complex $z$ plane.}
\label{fig:zplaneB14IH}
\end{center}
\end{figure}

Next, we note that $R_{23}$ can take values between $0$ and $1$.
It vanishes in the case of an exact two-zero texture and becomes
maximal in the absence of any hierarchy among $\hat{\kappa}_2$
and $\hat{\kappa}_3$.
In this sense, the zeros of $R_{23}$ in the complex $z$ plane
indicate the presence
of an exact two-zero texture in the neutrino Yukawa matrix.
Likewise, particularly small values of $R_{23}$ signal a strong
separation between the two smallest and the four largest Yukawa
couplings.
We demonstrate this behavior of $R_{23}$ in the complex $z$ plane
in Fig.~\ref{fig:zplaneB14IH}, where we plot $R_{23}$ as a function of $z$
for two different pairs of $\left(\delta,\sigma\right)$ values
that allow to realize a $B_{1,4}$ or a $B_{2,5}$ texture, respectively.


Fig.~\ref{fig:zplaneB14IH} illustrates that $R_{23}$ is periodic
in the real component of $z$ with a period of $\pi/2$,
\begin{align}
R_{23}\left(\delta,\sigma;z\right) =
R_{23}\left(\delta,\sigma;z+n\,\frac{\pi}{2}\right) \,, \quad n \in \mathbb{Z} \,.
\end{align}
This can be understood by noting that a shift of $z$
by an amount $\pi/2$ along the real axis basically corresponds to
nothing but a column exchange in the Yukawa matrix $\kappa_{\alpha I}$
(see Eq.~\eqref{eq:CIpara0}),
\begin{align}
z \rightarrow z + \frac{\pi}{2} \quad\Rightarrow\quad
\kappa_{\alpha 1} \rightarrow \kappa_{\alpha 2} \,, \quad
\kappa_{\alpha 2} \rightarrow - \kappa_{\alpha 1} \,.
\end{align}
Up to possible sign changes, this transformation, therefore, leaves the list
of sorted Yukawa couplings, $\hat{\kappa}$, unaffected, so
that the value of $R_{23}$ remains the same.
Similarly, it is worthwhile pointing out that $R_{23}$
as a function of $\delta$, $\sigma$, and $z$
is invariant under the following transformation,
\begin{align}
\delta \rightarrow -\delta \,, \quad
\sigma \rightarrow -\sigma \,, \quad
z \rightarrow z^* \,,
\end{align}
which amounts to a complex conjugation of the Yukawa couplings $\kappa_{\alpha I}$
(see Eq.~\eqref{eq:CIpara0}).
As the hierarchy parameter $R_{23}$ is defined in terms
of the \textit{absolute values} of the sorted Yukawa couplings,
$R_{23} = \left|\hat{\kappa}_2\right|/\left|\hat{\kappa}_3\right|$,
it remains unchanged under this transformation.
An important consequence of this property of $R_{23}$ is that
it essentially halves the domain of $R_{23}$ on the $\delta$ axis,
\begin{align}
\label{eq:ComplxCon}
R_{23}\left(2\pi-\delta,\sigma;z\right) =
R_{23}\left(\delta,\pi-\sigma;z^*\right) \,.
\end{align}
We note that it is this ``reflection symmetry'' of the parameter $R_{23}$
in the $\left(\delta,\sigma\right)$ plane, which is responsible for the fact that,
in Eq.~\eqref{eq:dsB1245}, we actually find \textit{two} solutions for
$\delta$ and $\sigma$ for both two-zero textures.
In studying $R_{23}$ as a function of $\delta$, $\sigma$, and $z$, 
we, therefore, do not need to consider the full $\delta$ range, $\delta \in \left[0,2\pi\right)$,
but can restrict ourselves to the restricted interval $\delta \in \left[0,\pi\right]$.


\begin{figure}
\begin{center}
\includegraphics[height=0.43\textheight]{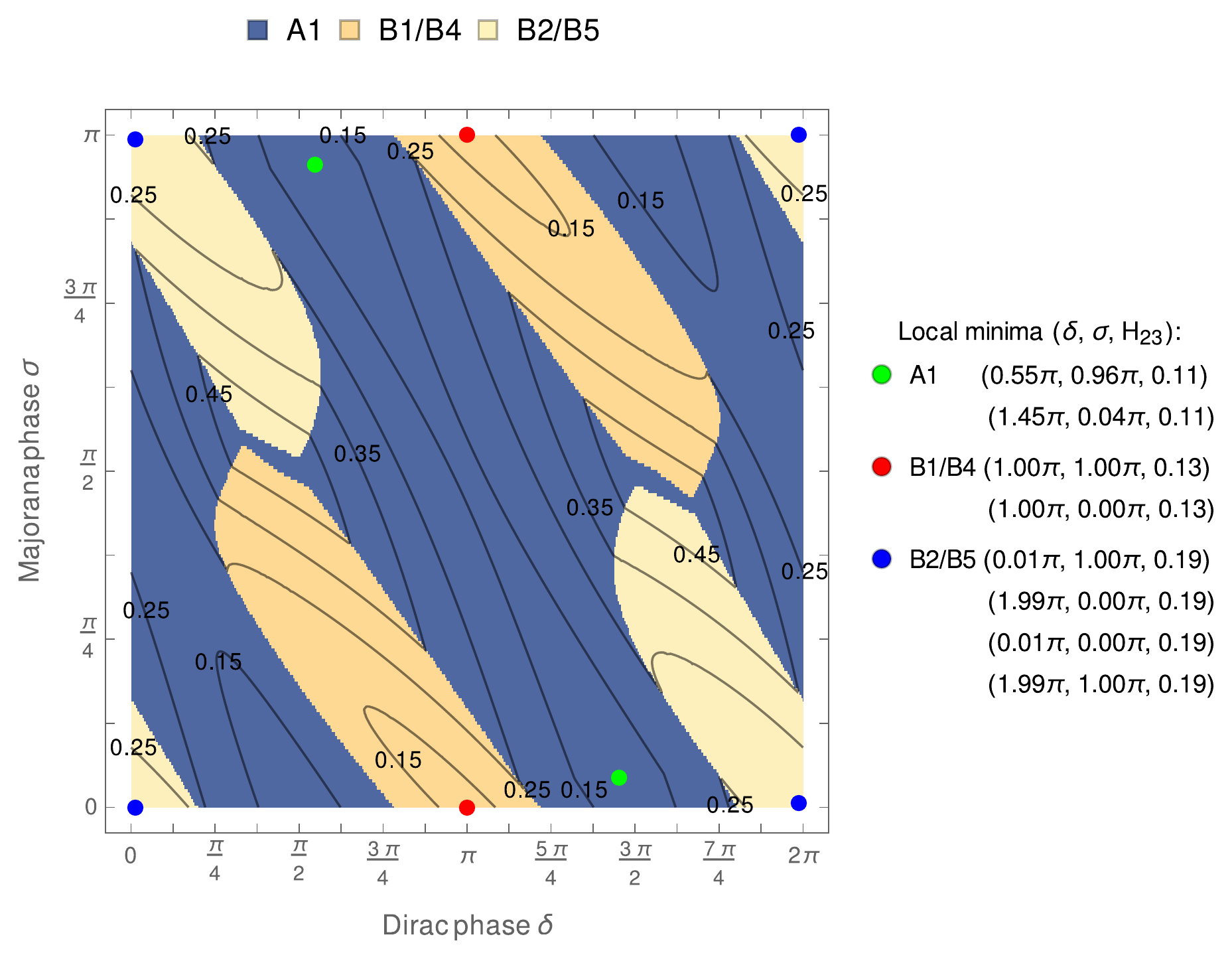}

\bigskip
\includegraphics[height=0.43\textheight]{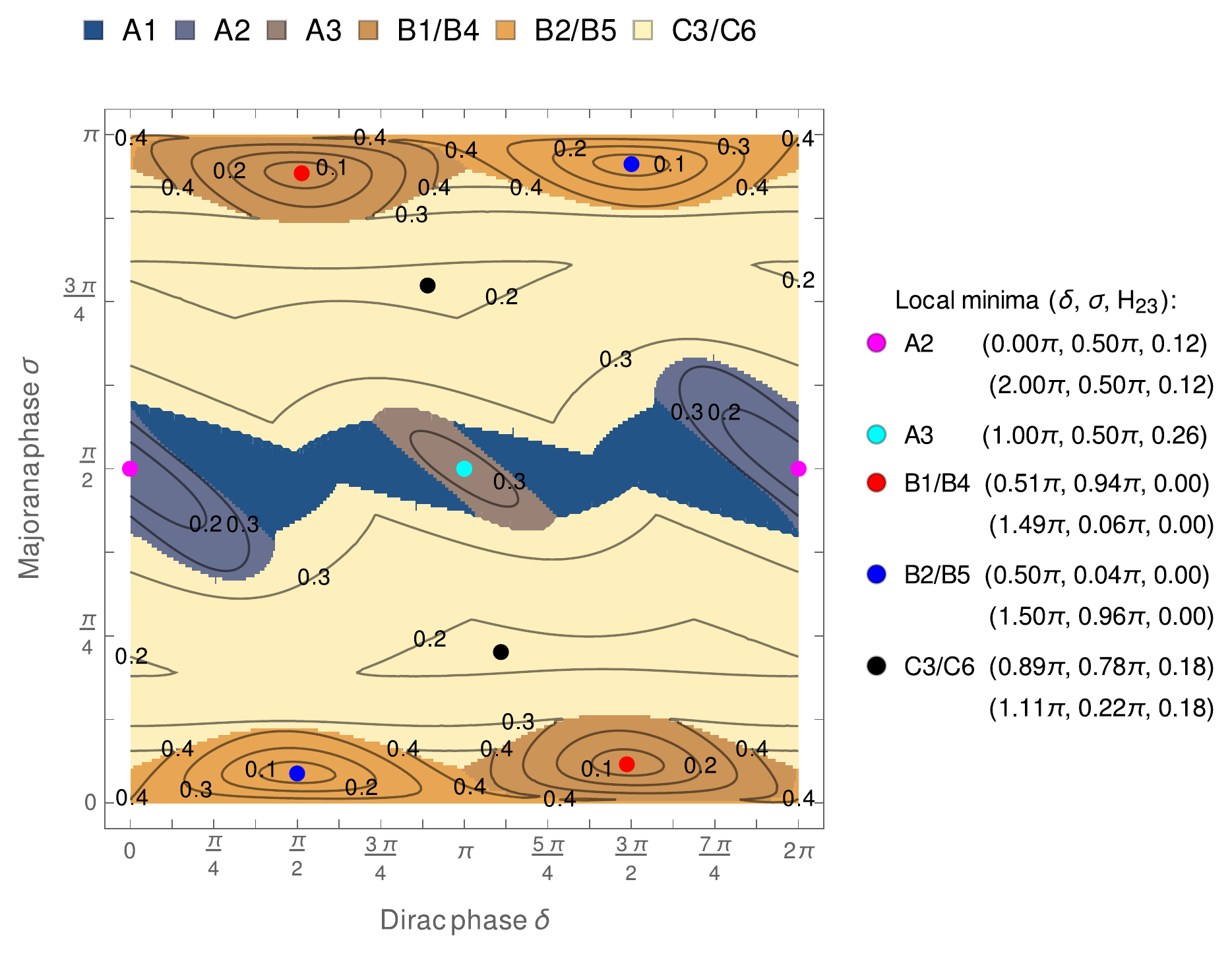}
\caption{Minimized hierarchy parameter $H_{23}$ as a function of the $CP$-violating
phases $\delta$ and $\sigma$ for normal \textbf{(upper panel)} and inverted neutrino
mass ordering \textbf{(lower panel)}, see Eq.~\eqref{eq:HierParam}.
The contour lines represent the function values of $H_{23}$ in the
$\left(\delta,\sigma\right)$ plane, while the color shading indicates the corresponding
(approximate) Yukawa textures according to Eq.~\eqref{eq:ABC}.
The colored points mark the local minima of $H_{23}$, see also Tab.~\ref{tab:results}.}
\label{fig:H23Scan}
\end{center}
\end{figure}


Finally, it is interesting to study $R_{23}$ in the limit of large
$\left|\textrm{Im}\left\{z\right\}\right|$.
Unlike the real component of $z$, the imaginary component of $z$
is not restricted to a periodic interval.
$\left|\textrm{Im}\left\{z\right\}\right|$ can, therefore, take
arbitrarily large values, which is the reason why
the parameter space of our model is unbounded.
Let us first consider the individual rescaled Yukawa couplings $\kappa_{\alpha I}$
in the limit of large $\left|\textrm{Im}\left\{z\right\}\right|$.
Decomposing the complex parameter $z$ into its real
and imaginary parts, $z = z_R + i\, z_I$, the Casas-Ibarra parametrization
in Eq.~\eqref{eq:CIpara0} may be rewritten as follows,
\begin{align}
\label{eq:CIparazRI}
\kappa_{\alpha 1} & = \frac{1}{2} \left[
\left(V_{\alpha k} + i\,V_{\alpha l}\right)e^{z_I-i\,z_R} + 
\left(V_{\alpha k} - i\,V_{\alpha l}\right)e^{-z_I+i\,z_R}
\right]
\,, \\ \nonumber
\kappa_{\alpha 2} & = \frac{1}{2}\left[
\left(V_{\alpha l} - i\,V_{\alpha k}\right)e^{z_I-i\,z_R} + 
\left(V_{\alpha l} + i\,V_{\alpha k}\right)e^{-z_I+i\,z_R}
\right] \,,
\end{align}
where $\left(k,l\right) = \left(2,3\right)$ in the NH case
and $\left(k,l\right) = \left(1,2\right)$ in the IH case.
In the limit of large $\left|z_I\right|$, we can either neglect
the terms proportional to $e^{z_I}$ or to $e^{-z_I}$, such that
Eq.~\eqref{eq:CIparazRI} simplifies to
\begin{align}
\label{eq:kappazI}
z_I \gg 1 & \quad\Rightarrow\quad \kappa_{\alpha 1} \simeq i\, \kappa_{\alpha 2} \simeq
\frac{1}{2}\left(V_{\alpha k} + i\,V_{\alpha l}\right)e^{z_I-i\,z_R} \,, \\ \nonumber
z_I \ll -1 & \quad\Rightarrow\quad \kappa_{\alpha 2} \simeq i\, \kappa_{\alpha 1} \simeq
\frac{1}{2}\left(V_{\alpha l} + i\,V_{\alpha k}\right)e^{-z_I+i\,z_R} \,.
\end{align}
In either case, we find that the two columns in $\kappa_{\alpha I}$ become---up
to a relative phase---equal to each other.
For $\left|z_I\right| \gg 1$, the list of sorted couplings, $\hat{\kappa}$,
consequently takes the following form,
\begin{align}
\left|z_I\right| \gg 1 & \quad\Rightarrow\quad 
\hat{\kappa} \simeq \left(
\kappa_{\alpha_11},\kappa_{\alpha_12},
\kappa_{\alpha_21},\kappa_{\alpha_22},
\kappa_{\alpha_31},\kappa_{\alpha_32}\right) \,,
\end{align}
where the indices $\left(\alpha_1,\alpha_2,\alpha_3\right)$ represent the
permutation of $\left(e,\mu,\tau\right)$ that satisfies
the relation $\left|\kappa_{\alpha_1I}\right| \leq \left|\kappa_{\alpha_2I}\right|\leq
\left|\kappa_{\alpha_3I}\right|$.
Physically, such a Yukawa configuration corresponds to an (approximate)
alignment in flavor space of the two linear combinations of lepton flavors
that couple to $N_1$ and $N_2$, respectively.
Without any additional physical assumption, e.g., an additional flavor symmetry
at high energies, such an alignment, however, appears unnatural or fine-tuned.
On top of that, too close an alignment results in a flavor structure that only
manages to account for the low-energy data on neutrino masses and mixings
by means of precise cancellations among rather large Yukawa couplings---which
also appears unnatural.
In the limit of exact flavor alignment, $\kappa_{\alpha1}\equiv\kappa_{\alpha2}$,
it is even impossible to explain all of
the low-energy data, because the number of available high-energy DOFs is simply too small.
For these reasons, we will be more interested in values of $\left|z_I\right|$ of
$\mathcal{O}(1)$ in the following.
On the other hand, as we cannot exclude the possibility of aligned lepton
flavors, we will, of course, keep our analysis as general
as possible and also devote some attention
to the case of large $\left|z_I\right|$ in our parameter scan.%
\footnote{See also the recent analysis in Ref.~\cite{Rink:2016knw}, which regards
the issue of flavor alignment as a feature, not a bug.
In this model, the assumption of approximate flavor alignment
in the neutrino Yukawa matrix is combined with a certain UV
flavor model of the Froggatt-Nielsen type~\cite{Froggatt:1978nt}.
Together, these model assumptions result in a truly minimal and highly predictive
realization of the type-I  seesaw mechanism with only two right-handed neutrinos.}


An important implication of Eq.~\eqref{eq:kappazI} for our purposes is that, in
the regime of large $\left|z_I\right|$, ratios of Yukawa couplings no longer
depend on $z$.
This means that, for $z_I \rightarrow \pm \infty$, the hierarchy parameter
$R_{23}$ is bound to asymptotically approach a constant value, $R_{23}^+$
and $R_{23}^-$, respectively,
\begin{align}
z_I \rightarrow \pm \infty \quad\Rightarrow\quad
R_{23} \rightarrow R_{23}^\pm =
\left|\frac{V_{\alpha_1 k} \pm i\,V_{\alpha_1 l}}
{V_{\alpha_2 k} \pm i\,V_{\alpha_2 l}}\right| \,.
\label{eq:R23pm}
\end{align}
In the case of the two scenarios shown in Fig.~\ref{fig:zplaneB14IH},
we obtain, e.g., the following asymptotic values,
\begin{align}
\left(R_{23}^+,R_{23}^-\right) = \begin{cases}
\left(0.81, 0.68\right) & ; \quad \left(\delta,\sigma\right) = \left(0.51,0.94\right) \pi \\
\left(0.61, 0.89\right) & ; \quad \left(\delta,\sigma\right) = \left(0.50,0.04\right) \pi
\end{cases} \,,
\end{align}
which are all more or less close to $1$.
This is to say that, in these two scenarios, we are not missing any strong hierarchy
in the neutrino Yukawa matrix that could potentially be realized in regions of the complex
$z$ plane that are not displayed in Fig.~\ref{fig:zplaneB14IH}.
Instead, the local minima depicted in Fig.~\ref{fig:zplaneB14IH} are really
the only minima of $R_{23}$ there are in the complex $z$ plane.


\section{Maximal Yukawa hierarchies in dependence of
\texorpdfstring{\boldmath{$\delta$}}{delta} and
\texorpdfstring{\boldmath{$\sigma$}}{sigma}}
\label{sec:scan}


\begin{table}
\begin{center}
\begin{tabular}{|c||ccccc|}\hline
 &
$\left(\delta,\sigma\right)/ \pi$ & Texture & $H_{23}$ & $z_I$ &
Rescaled Yukawa matrix $\kappa\times 10^7$ \\\hline\hline
NH-1 & \specialcell[c]{$\left(0.55,0.96\right)$\\$\left(1.45,0.04\right)$}
& A1 & $0.11$ & $\ll -1$ &
$\begin{pmatrix}
0.41\, e^{1.02\,i\pi} & 0.41\, e^{1.52\,i\pi} \\
3.85\, e^{1.38\,i\pi} & 3.85\, e^{1.88\,i\pi} \\
3.90\, e^{1.60\,i\pi} & 3.90\, e^{0.10\,i\pi} 
\end{pmatrix} f\left(z\right)$ 
\vphantom{$\bigg(_{\bigg(_{\big(}}^{\bigg(^{\big(}}$}
\\\hline
NH-2 & $\left(1.00,0.00\right)$ & B1/B4 & $0.13$ & $\mathcal{O}(1)$ &
$\begin{pmatrix}
           -0.18 &  \phantom{-} 1.42\,\phantom{i} \\
\phantom{-} 3.83 &             -0.18\,\phantom{i} \\
\phantom{-} 3.11 &             -2.73\,\phantom{i}  
\end{pmatrix} \phantom{f\left(z\right)}$ 
\vphantom{$\bigg(_{\bigg(_{\big(}}^{\bigg(^{\big(}}$}
\\\hline
NH-3 & \specialcell[c]{$\left(0.01,0.00\right)$\\$\left(1.99,0.00\right)$}
& B2/B5 & $0.19$ & $\mathcal{O}(1)$ &
$\begin{pmatrix}
0.27\,e^{0.01\,i\pi} & 1.41\,e^{0.01\,i\pi} \\
2.76\,\phantom{e^{0.01\,i\pi}} & 2.53\,\phantom{e^{0.01\,i\pi}} \\
4.21\,\phantom{e^{0.01\,i\pi}} & 0.27\,\phantom{e^{0.01\,i\pi}}
\end{pmatrix} \phantom{f\left(z\right)}$ 
\vphantom{$\bigg(_{\bigg(_{\big(}}^{\bigg(^{\big(}}$}
\\\hline
\hline
IH-1 & $\left(0.00,0.50\right)$ & A2 & $0.12$ & $\gg +1$ & 
$\begin{pmatrix}
\phantom{-}7.30 & -7.30\,i \\
\phantom{-}0.22 & -0.22\,i \\
-1.92 & \phantom{-}1.92\,i 
\end{pmatrix} f\left(z\right)$ 
\vphantom{$\bigg(_{\bigg(_{\big(}}^{\bigg(^{\big(}}$}
\\\hline
IH-2 & $\left(1.00,0.50\right)$ & A3 & $0.26$ & $\gg +1$ &
$\begin{pmatrix}
\phantom{-}7.30 & -7.30\,i \\
\phantom{-}1.87 & -1.87\,i \\
-0.49 & \phantom{-}0.49\,i 
\end{pmatrix} f\left(z\right)$ 
\vphantom{$\bigg(_{\bigg(_{\big(}}^{\bigg(^{\big(}}$}
\\\hline
IH-3 & \specialcell[c]{$\left(0.51,0.94\right)$\\ $\left(1.49,0.06\right)$}
& B1/B4 & $0.00$ & $\mathcal{O}(1)$ &
$\begin{pmatrix}
0 & 5.24\, e^{1.02\,i\pi} \\
3.54\, e^{1.04\,i\pi} & 0 \\
4.06\, e^{0.04\,i\pi} & 1.19\, e^{0.53\,i\pi} 
\end{pmatrix} \phantom{f\left(z\right)}$ 
\vphantom{$\bigg(_{\bigg(_{\big(}}^{\bigg(^{\big(}}$}
\\\hline
IH-4 & \specialcell[c]{$\left(0.50,0.04\right)$\\ $\left(1.50,0.96\right)$}
& B2/B5 & $0.00$ & $\mathcal{O}(1)$ &
$\begin{pmatrix}
0 & 5.25\, e^{1.99\,i\pi} \\
3.52\, e^{0.97\,i\pi} & 1.04\, e^{1.49\,i\pi} \\
4.05\, e^{1.97\,i\pi} & 0
\end{pmatrix} \phantom{f\left(z\right)}$ 
\vphantom{$\bigg(_{\bigg(_{\big(}}^{\bigg(^{\big(}}$}
\\\hline
IH-5 & \specialcell[x]{$\left(0.89,0.78\right)$\\ $\left(1.11,0.22\right)$}
& C3/C6 & $0.18$ & $\mathcal{O}(1)$ &
$\begin{pmatrix}
5.83\, e^{1.06\,i\pi} & 3.42\, e^{0.55\,i\pi}\\
0.62\, e^{0.95\,i\pi} & 3.42\, e^{0.19\,i\pi}\\
0.62\, e^{0.94\,i\pi} & 3.42\, e^{1.14\,i\pi}
\end{pmatrix} \phantom{f\left(z\right)}$ 
\vphantom{$\bigg(_{\bigg(_{\big(}}^{\bigg(^{\big(}}$}
\\\hline
\hline
\end{tabular}
\caption{Minima of the minimized hierarchy parameter $H_{23}$ in the
$\left(\delta,\sigma\right)$ plane for normal and inverse light-neutrino mass
ordering, respectively, see Eq.~\eqref{eq:HierParam}.
Nomenclature for the (approximate) textures according to Eq.~\eqref{eq:ABC}.
Large values of $\left|z_I\right|$ correspond to \textit{flavor alignment} in the neutrino
Yukawa matrix $\kappa_{\alpha I}$, while $\mathcal{O}(1)$ values of $\left|z_I\right|$
do not imply any particular correlation among the two columns in $\kappa_{\alpha I}$,
see Eq.~\eqref{eq:CIparazRI}.
At large values of $\left|z_I\right|$, all Yukawa couplings scale with $f\left(z\right) = 1/2\,
\exp\left[\textrm{sgn}\left\{z_I\right\}\left(z_I - i\,z_R\right)\right]$,
see Eq.~\eqref{eq:kappazI}.}
\label{tab:results}
\end{center}
\end{table}


\begin{figure}
\begin{center}
\includegraphics[height=0.43\textheight]{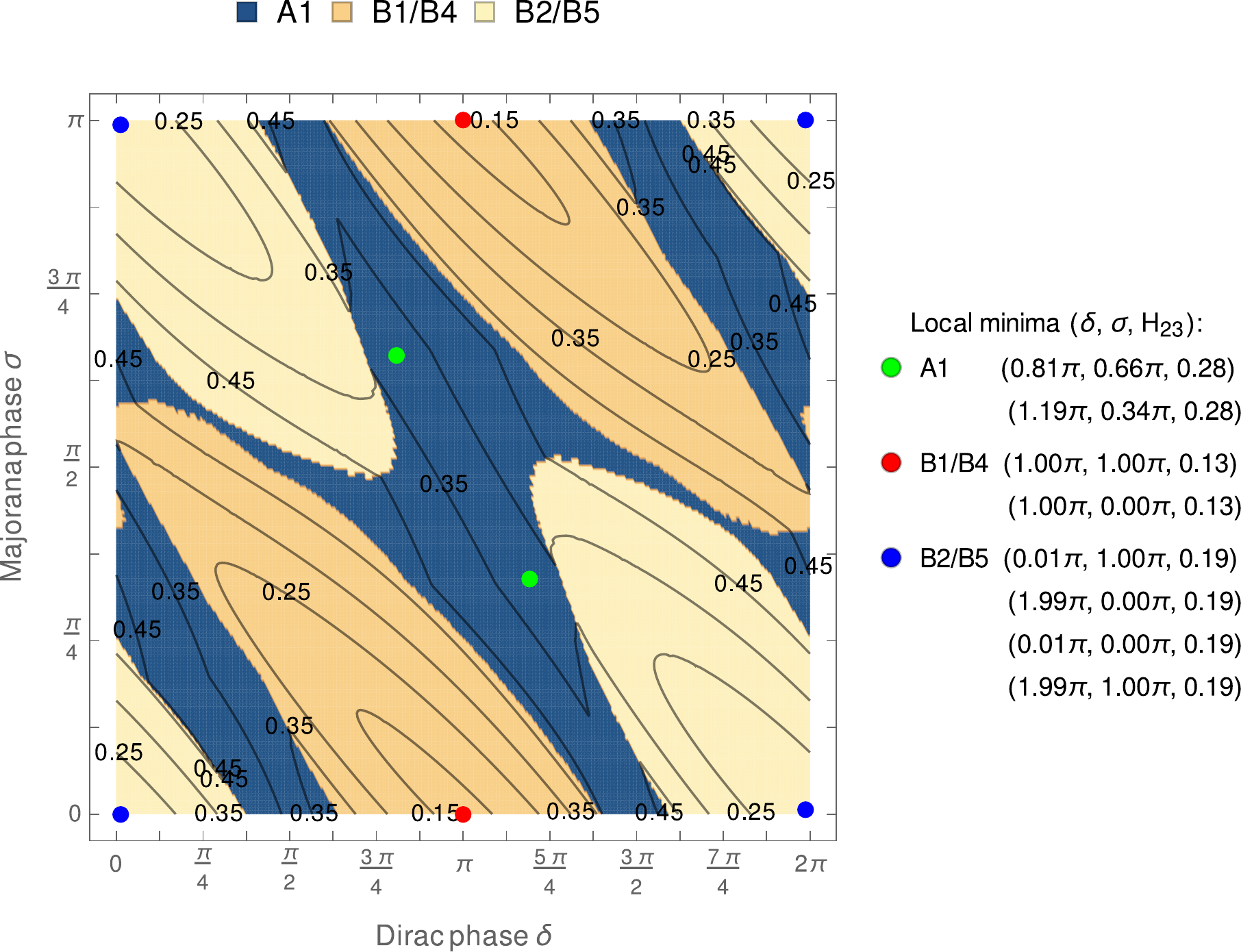}

\bigskip
\includegraphics[height=0.43\textheight]{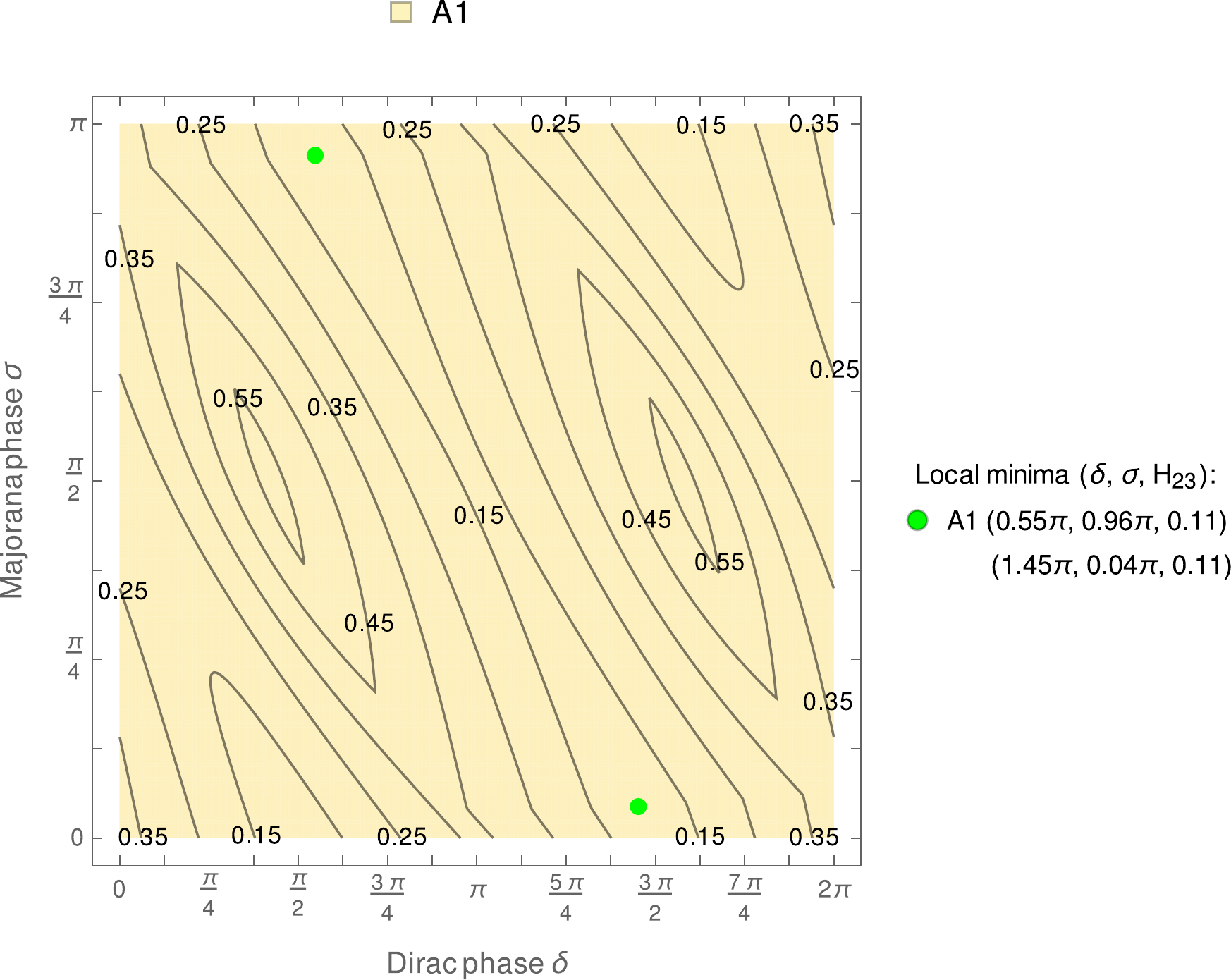}
\caption{Minimized hierarchy parameter $H_{23}$ as a function of the $CP$-violating
phases $\delta$ and $\sigma$ for \textbf{normal mass ordering}, see Eq.~\eqref{eq:HierParam}.
The \textbf{upper panel} displays the results of minimizing $R_{23}$ only over values of
$\left|z_I\right|$ of $\mathcal{O}(1)$ (no flavor alignment), while in the
\textbf{lower panel}, we plot the asymptotic value $\min\left\{R_{23}^-,R_{23}^+\right\}$
at $\left|z_I\right| \gg 1$ (flavor alignment), see Eq.~\eqref{eq:R23pm}.
Contour lines, color shading, colored points, and plot legends as in Fig.~\ref{fig:H23Scan}.}
\label{fig:NHsep} 
\end{center}
\end{figure}


As explained in the previous section, an approximate [exact] two-zero texture in a
given neutrino Yukawa matrix is characterized by a particularly small value [zero]
of the hierarchy parameter $R_{23}\left(\delta,\sigma;z\right)$.
For given values of the $CP$-violating phases $\delta$ and $\sigma$,
we would, therefore, like to know how small a value $R_{23}$ can possibly
take as a function of the rotation angle $z$.
To answer this question, we are now going to study the \textit{minimized hierarchy parameter}
$H_{23}\left(\delta,\sigma\right)$,
\begin{align}
\label{eq:HierParam}
H_{23}\left(\delta,\sigma\right) = \min_{z_R}  \min_{z_I} R_{23}\left(\delta,\sigma;z\right) \,.
\end{align}
That is, for each pair of $\left(\delta,\sigma\right)$ values, we are going
to minimize the ratio of sorted Yukawa couplings $R_{23}$
over the two real DOFs that are inaccessible at low energies in our model, $z_R$ and $z_I$.
This eliminates the dependence of $R_{23}$ on the ``unphysical'' parameter $z$
and leaves us with a well-defined function in the $\left(\delta,\sigma\right)$ plane:
At any given point in this plane, the function $H_{23}$ tells us how good a hierarchy
in the neutrino Yukawa matrix one could, in principle, obtain.
Without any extra assumptions, there is, of course, no reason why a certain value
of $H_{23}$ should, indeed, be realized in Nature.
It is always possible that,
instead of the maximal Yukawa hierarchy for a particular pair of $\left(\delta,\sigma\right)$
values, only a weaker hierarchy is actually realized
(which would then correspond to a point in the complex $z$ plane away from the
global minimum of $R_{23}$).
In this sense, our analysis will provide us with model-independent
\textit{upper bounds} on the quality of an approximate two-zero
texture in dependence of $\delta$ and $\sigma$ that any UV completion
has to satisfy.
Here, note that a complete reconstruction of the entire neutrino Yukawa matrix from
a measurement of $\delta$ and $\sigma$ is not feasible in our framework, anyway.
The advantage of our approach, however, is that it will provide us with
meaningful constraints on UV completions of the minimal seesaw model,
no matter what the experimental outcome for $\delta$ and $\sigma$ will be.


It is straightforward to extend the philosophy behind our approach to other,
more complicated neutrino mass models as well as to arbitrary textures.
In the end, one always only has to follow the same three steps.
(i) Express all high-energy Yukawa couplings in terms of two sets of parameters:
low-energy observables on the one hand and ``unphysical'' parameters that are
inaccessible at low energies on the other hand;
(ii) define an appropriate ratio of (sorted) Yukawa couplings that is relevant to a
certain texture; and
(iii) marginalize this ratio over all unobservable parameters.
This procedure should always results in phenomenological constraints
on the Yukawa structures of interest in dependence of the available observables
at low energies.
Of course, the technical complexity of this algorithm grows with the number of
inaccessible parameters.
From this perspective, our analysis in this paper benefits from the fact
that we are only dealing with two auxiliary parameters, $z_R$ and $z_I$.
Or put differently:
While our method appears to be applicable also in a more general context,
we choose to illustrate our approach in the context of the minimal seesaw neutrino
mass model because of its sheer simplicity.


A practical obstacle that one encounters when attempting to compute $H_{23}$
as a function of $\delta$ and $\sigma$ is that the hierarchy parameter $R_{23}$
is not a smooth function of the complex rotation angle $z$.
Whenever the ordering among the sorted Yukawa couplings changes in consequence of a
variation of $z$, $R_{23}$ experiences a ``kink'' at which it is longer differentiable.
This observation leads us to forgo any further analytical calculations in the
following.
Instead, we shall search for the global minima of $R_{23}$ in the complex $z$ plane
by means of a fully numerical analysis.

 
As discussed in Sec.~\ref{sec:hierarchy}, large values of $\left|z_I\right|$
lead to an alignment of the two columns in the neutrino Yukawa matrix,
see Eq.~\eqref{eq:kappazI}.
Physically, such a flavor alignment signals the presence of some kind of
flavor symmetry at high energies, for which we do not have a proper justification from a
low-energy perspective.
On the other hand, we also cannot \textit{a priori} exclude the possibility
of flavor alignment.
In the following, we are, therefore, going to explicitly distinguish between
these two cases, i.e., between Yukawa matrices that do exhibit flavor alignment,
$\left|z_I\right| \gg 1$, and those that do not,
$\left|z_I\right| \sim \mathcal{O}(1)$.
In this way, we will be able to track under which conditions certain textures arise 
and whether they rely on the assumption of flavor alignment or not.


\begin{figure}
\begin{center}
\includegraphics[height=0.43\textheight]{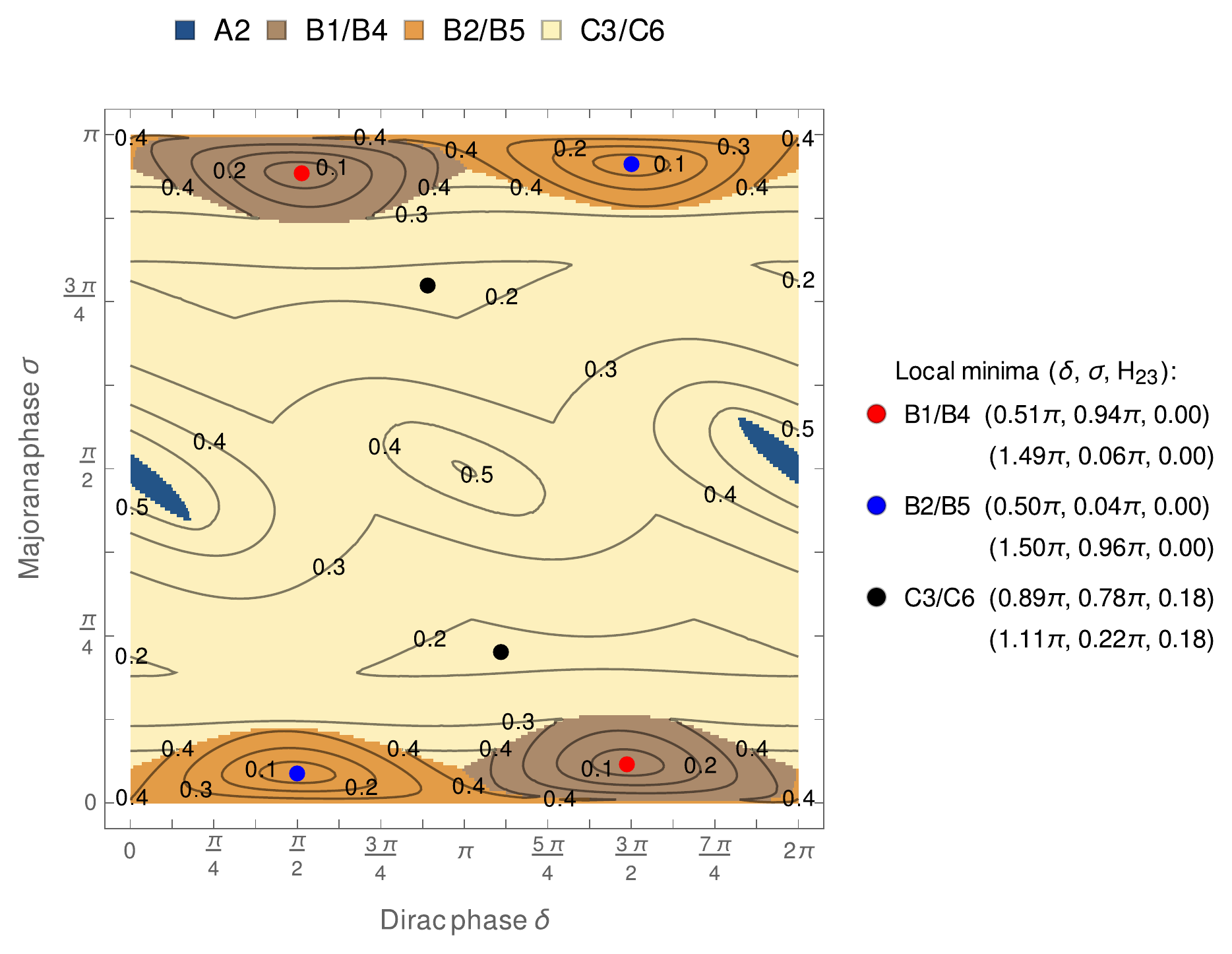}

\bigskip
\includegraphics[height=0.43\textheight]{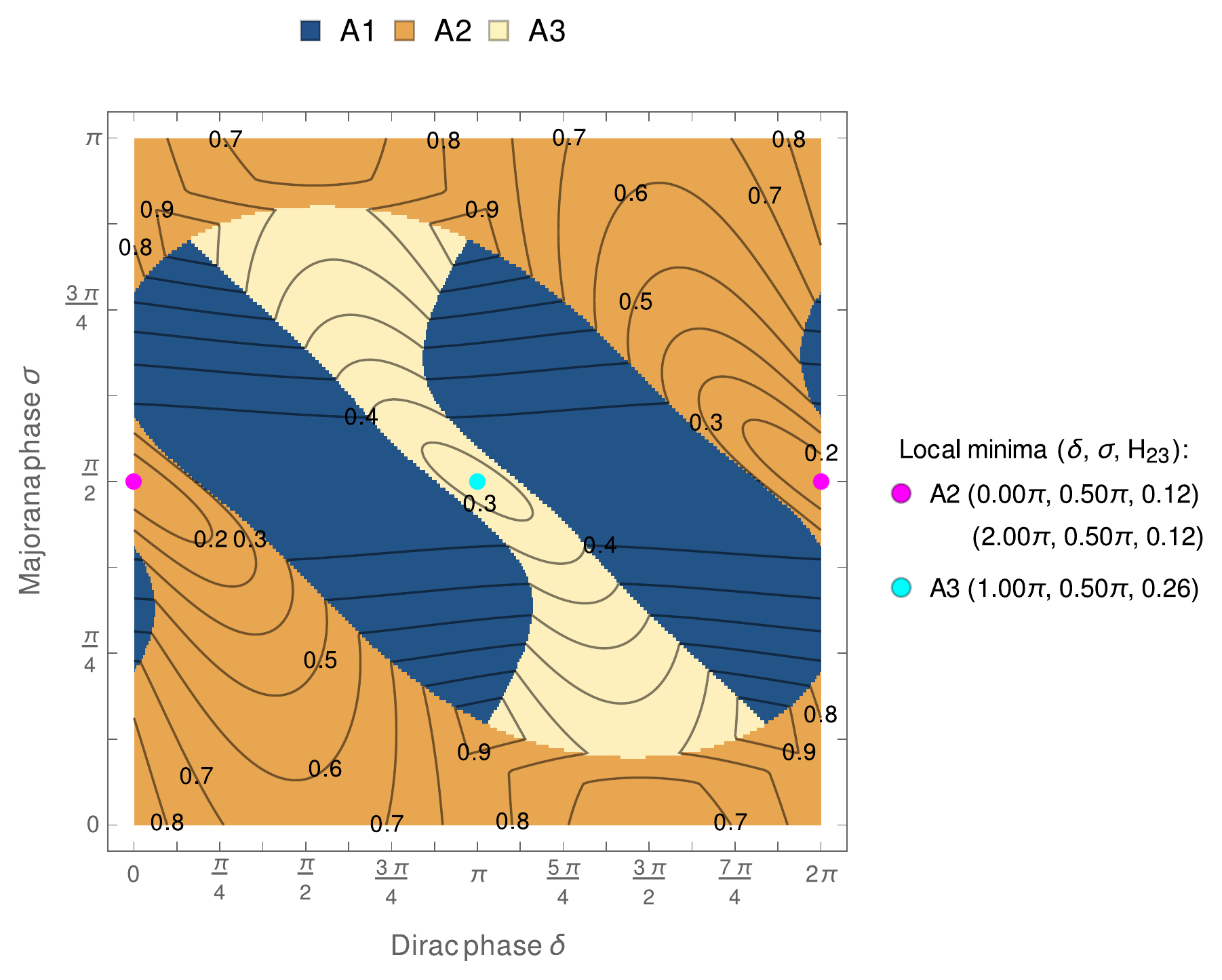}
\caption{Minimized hierarchy parameter $H_{23}$ as a function of the $CP$-violating
phases $\delta$ and $\sigma$ for \textbf{inverted mass ordering}, see Eq.~\eqref{eq:HierParam}.
The \textbf{upper panel} displays the results of minimizing $R_{23}$ only over values of
$\left|z_I\right|$ of $\mathcal{O}(1)$ (no flavor alignment), while in the
\textbf{lower panel}, we plot the asymptotic value $\min\left\{R_{23}^-,R_{23}^+\right\}$
at $\left|z_I\right| \gg 1$ (flavor alignment), see Eq.~\eqref{eq:R23pm}.
Contour lines, color shading, colored points, and plot legends as in Fig.~\ref{fig:H23Scan}.}
\label{fig:IHsep} 
\end{center}
\end{figure}

\begin{figure}[t!]
\begin{center}
\includegraphics[height=0.465\textwidth]{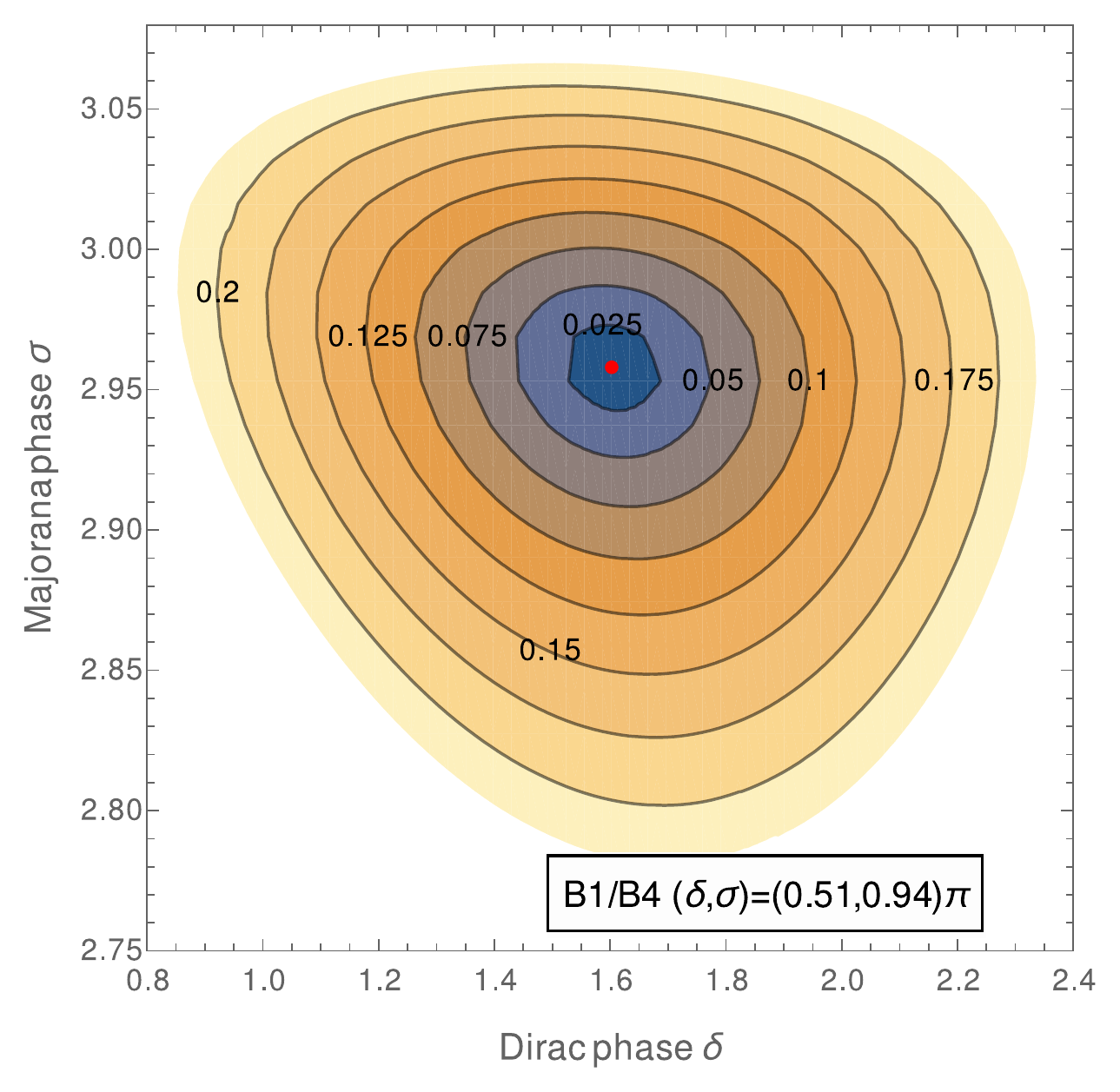}\hfill
\includegraphics[height=0.465\textwidth]{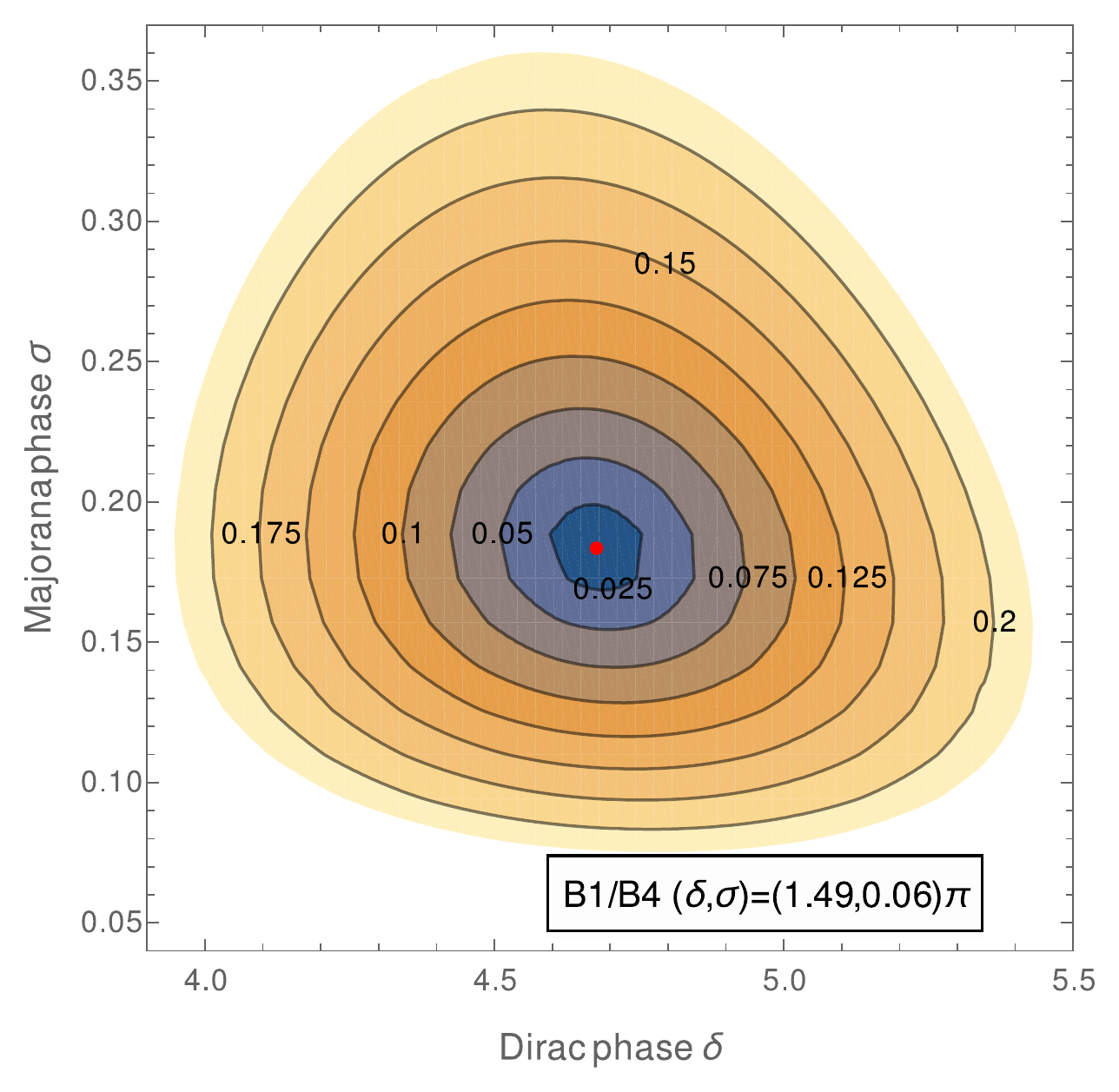}

\bigskip
\includegraphics[height=0.465\textwidth]{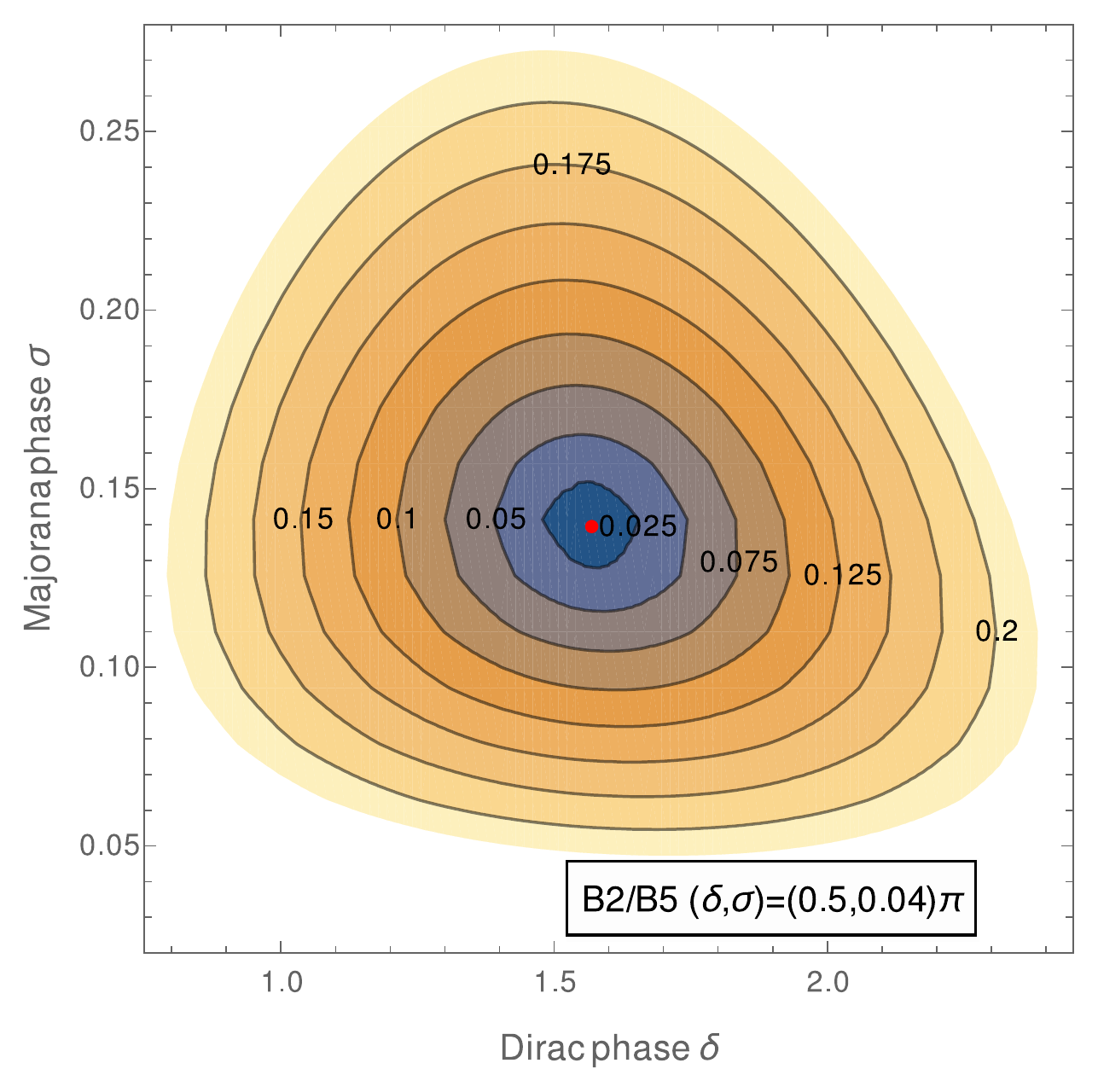}\hfill
\includegraphics[height=0.465\textwidth]{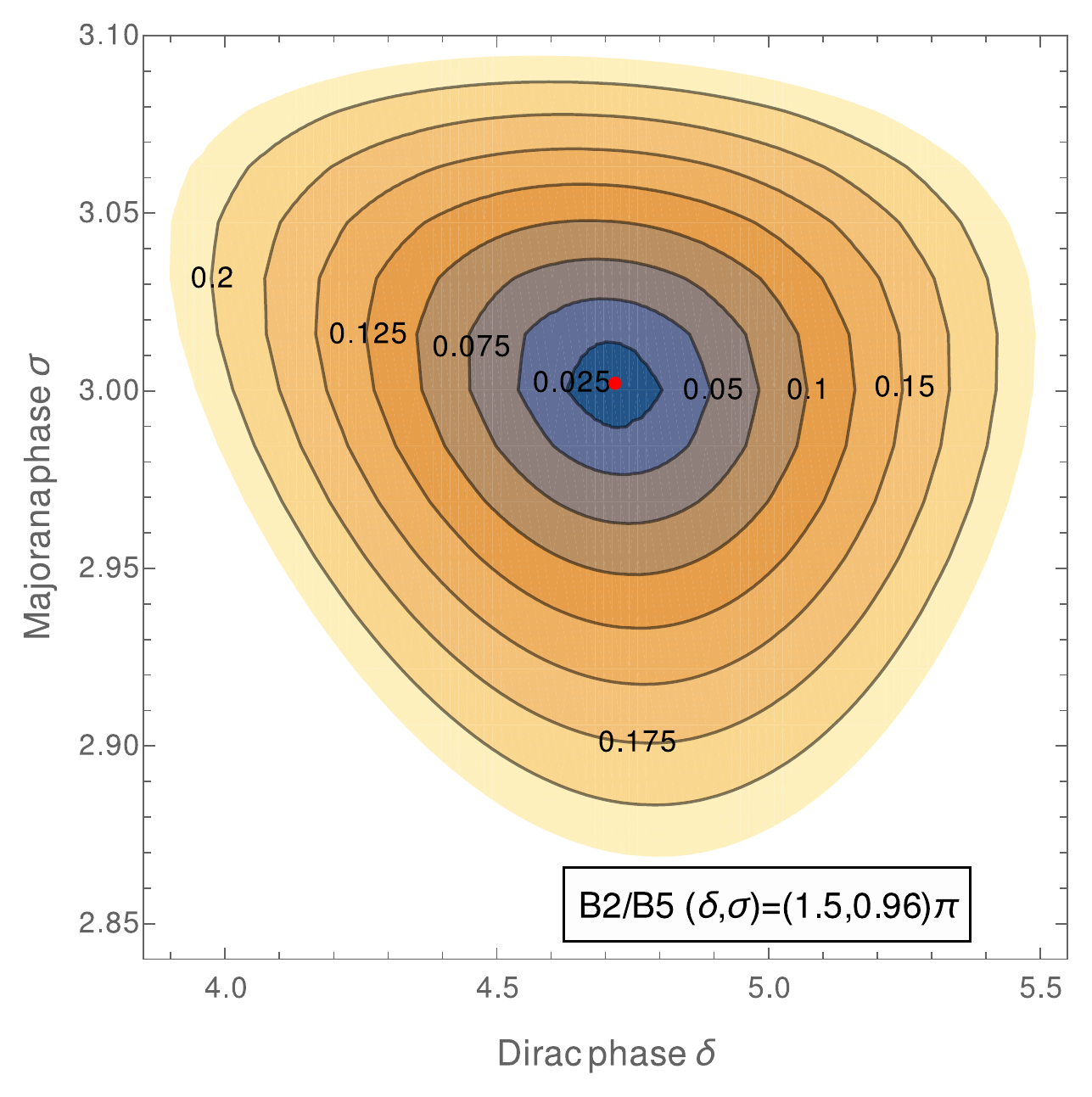}
\caption{Minimized hierarchy parameter $H_{23}$ as a function of 
$\delta$ and $\sigma$ in the vicinity of its exact zeros (in the case of
an inverted mass ordering), see Eq.~\eqref{eq:dsB1245}.
For both possible two-zero textures, $B_{1,4}$ and $B_{2,5}$, we respectively find two
zeros in the $\left(\delta,\sigma\right)$ plane, which are related to each other
via the reflection symmetry in Eq.~\eqref{eq:ComplxCon}.}
\label{fig:zoom} 
\end{center}
\end{figure}

The outcome of our numerical scan is shown in Figs.~\ref{fig:H23Scan}, \ref{fig:NHsep},
\ref{fig:IHsep}, and \ref{fig:zoom}.
In Fig.~\ref{fig:H23Scan}, we present our results for $H_{23}$ as a function of
$\delta$ and $\sigma$ after minimizing $R_{23}$ over the \textit{entire}
complex $z$ plane.
The two panels of Fig.~\ref{fig:H23Scan} correspond to NH and IH, respectively.
Figs.~\ref{fig:NHsep} and \ref{fig:IHsep} illustrate how these results
decompose into contributions from ``flavor-aligned'' ($\left|z_I\right| \gg 1$)
and ``flavor-nonaligned'' Yukawa matrices ($\left|z_I\right| \sim \mathcal{O}(1)$).
Here, Fig.~\ref{fig:NHsep} summarizes our results for NH,
while Fig.~\ref{fig:IHsep} summarizes our results for IH.
The upper panels of Figs.~\ref{fig:NHsep} and \ref{fig:IHsep} show
our respective results for $H_{23}$ after minimizing $R_{23}$ over $z$ values
not far from the real axis, $\left|z_I\right| \sim \mathcal{O}(1)$, whereas
in the lower panels of Figs.~\ref{fig:NHsep} and \ref{fig:IHsep},
we plot the respective minimum of the two asymptotic values at large
$\left|z_I\right|$, $\min\left\{R_{23}^-,R_{23}^+\right\}$, see Eq.~\eqref{eq:R23pm}.
In Fig.~\ref{fig:zoom}, we finally present a close-up of our results for $H_{23}$
in the vicinity of its exact zeros in the case of an inverted mass ordering.
A summary of all local minima of the function $H_{23}$ in the
$\left(\delta,\sigma\right)$ plane is given in Tab.~\ref{tab:results}.


Let us now comment on our numerical findings in more detail:
As shown in Refs.~\cite{Harigaya:2012bw,Zhang:2015tea}, the only \textit{exact}
two-zero textures that can be realized in the type-I seesaw model with two
right-handed neutrinos are textures $B_{1,4}$ and $B_{2,5}$,
and that only if one assumes an inverted hierarchy.
In our scan, we readily recover these two exact textures as zeros of $H_{23}$
in the $\left(\delta,\sigma\right)$ plane.
But in addition to that, our analysis has more to offer!
The conventional top-down approach merely leads
to precise predictions for $\delta$ and $\sigma$ under the
assumption of exact texture zeros.
If these predictions should turn out to be inconsistent with the experimental
data at some point (which is maybe not unlikely), exact two-zero textures would be ruled out
and the scenarios in Refs.~\cite{Harigaya:2012bw,Zhang:2015tea} would need to be
declared unviable.
In such a case, our analysis now shows how the assumption of exact texture zeros
would need to be relaxed, so as still accommodate the experimental data.
Fig.~\ref{fig:zoom} indicates how strong a hierarchy
in the neutrino Yukawa matrix we would \textit{at least} have
to assume, once $\delta$ and $\sigma$
should be measured at values that are slightly off the exact
predictions in Eq.~\eqref{eq:dsB1245}.
Shifts in the values of $\delta$ and $\sigma$ by, say, $0.1\,\pi$ compared
to the predictions in Eq.~\eqref{eq:dsB1245} would, e.g., require
$\mathcal{O}\left(10\,\%\right)$ corrections to an exact two-zero texture.
This generalization of the predictions in Eq.~\eqref{eq:dsB1245} is one of
our main results in this paper.
Our findings in Fig.~\ref{fig:zoom} correspond to the theoretical
error bars belonging to the ``central values'' in Eq.~\eqref{eq:dsB1245}.
In this sense, our study represents a crucial supplement to the analyses in
Refs.~\cite{Harigaya:2012bw,Zhang:2015tea},
which derive rather uncompromising predictions for specific and rigid
theoretical assumptions.


Moreover, we find that, in addition to $B_{1,4}$ and $B_{2,5}$ textures,
several other textures may be \textit{approximately} realized in our neutrino
mass model.
In the NH case, we encounter matrices that belong to Yukawa structures $A_1$,
$B_{1,4}$, and $B_{2,5}$; while in the IH case, we obtain structures
of type $A_1$, $A_2$, $A_3$, $B_{1,4}$, $B_{2,5}$, and $C_{3,6}$.%
\footnote{At this point, we merely classify all possible Yukawa matrices
according to the position of their two smallest elements (see Eq.~\eqref{eq:ABC}),
irrespectively of the corresponding value of the ratio $R_{23}$.
(Approximate) Yukawa \textit{textures} in the actual sense are only realized
if $R_{23}$ is particularly small, i.e.,
for $R_{23} \lesssim \mathcal{O}\left(0.1\right)$ or so.}
In certain regions in the $\left(\delta,\sigma\right)$ plane, i.e., in the vicinity
of the local minima of $H_{23}$, these structures partially exhibit strong hierarchies, 
so that they may be regarded as approximate textures
in the actual sense, see Tab.~\ref{tab:results}.
Here, an interesting observation is that, also in the case of
a normal mass ordering, the neutrino Yukawa matrix may exhibit
a strong hierarchy.
In the case of flavor alignment, we find, e.g., that at
$\left(\delta,\sigma\right) \simeq \left(0.55,0.96\right)\pi$
as well as at $\left(\delta,\sigma\right) \simeq \left(1.45,0.04\right)\pi$
an approximate $A_1$ texture can be realized, with the smallest two
entries in the neutrino Yukawa matrix being suppressed compared
to the next-largest entry by roughly a factor of $10$.
As soon as we relax the assumption of two \textit{exact}
texture zeros, the case of a normal mass hierarchy can, therefore, no longer
be excluded.
For sure, the type-I seesaw model with two right-handed neutrinos 
is still inconsistent with the assumption of an exact two-zero texture.
This conclusion does not change.
But a main result of our analysis is that also the NH case
becomes feasible again, once we allow for small
perturbations of $\mathcal{O}\left(10\,\%\right)$. 


Another interesting observation is that the minimized hierarchy
parameter $H_{23}$ turns out to have a preference for small function values close
to points in the $\left(\delta,\sigma\right)$ plane that conserve
$CP$ in the leptonic weak charged-current interactions
(see the discussion in Ref.~\cite{Agashe:2014kda} for more details),
\begin{align}
\frac{\left(\delta,\sigma\right)}{\pi} = \left(m,\frac{n}{2}\right) \,,
\quad m,n \in \mathbb{Z} \quad\Rightarrow\quad
\textrm{No $CP$ violation} \,.
\end{align}
This means that, out of the six ``new'' local minima of $H_{23}$ that we find in our analysis
(NH-1, NH-2, NH-3, IH-1, IH-2, IH-5), three minima (NH-2, IH-1, IH-2)
correspond to no $CP$ violation at all; in one minimum (NH-3), there
is barely no $CP$ violation; and only two ``new'' minima (NH-1, IH-5)
feature a sizable amount of $CP$ violation, see Tab.~\ref{tab:results}.
If the recent experimental hints for a ``maximal'' $CP$-violating phase
$\delta$ should be confirmed in upcoming experiments, the $CP$-conserving
scenarios NH-2, NH-3, IH-1, and IH-2 would, therefore, need to be abandoned. 


Last but not least, we comment on the interplay of ``flavor-aligned''
and ``flavor-nonaligned'' regions in Fig.~\ref{fig:H23Scan}.
In the NH case, the effect of extending our search to large values
of $\left|z_I\right|$ is that it (i) shrinks the region occupied
by $B_{1,4}$ and $B_{2,5}$ textures and (ii) shifts (as well as deepens)
the position of the local $A_1$ minima, see Fig.~\ref{fig:NHsep}.
In the IH case, ``flavor-aligned'' matrices are responsible
for the horizontal band of $A_1$, $A_2$, and $A_3$ structures at $\sigma \sim \frac{\pi}{2}$. 
Here, regions, in which the structures
$A_2$ and $A_3$ appear, exhibit local maxima
in the nonaligned case, see Fig.~\ref{fig:IHsep}.


\section{Stability under variations of the experimental input data}
\label{sec:stability}


In our numerical scan presented in Sec.~\ref{sec:scan}, we always employ
the best-fit values for all known experimental input values, see Tab.~\ref{tab:data}.
Our numerical results are, thus, subject to small variations in case the
``true'' values for the light-neutrino masses and mixing angles should eventually
be found to slightly deviate from the current best-fit values.
To get a feeling for this dependence of our results on the experimental data, we
are going to perform a naive $\chi^2$ analysis in the following.
In doing so, we will neglect the detailed likelihood functions in Ref.~\cite{Capozzi:2016rtj}
and simply treat all observables as if they were Gaussian.
It is then straightforward to define $\chi^2$,
\begin{align}
\chi^2\left(\left\{\mathcal{O}_i\right\}\right) =
\sum_{i=1}^5 \frac{\left(\mathcal{O}_i-\mathcal{O}_i^{\rm exp}\right)^2}
{\left(\Delta \mathcal{O}_i^{\rm exp}\right)^2} \,, \quad
\left\{\mathcal{O}_i\right\} = \left\{\delta m^2, \Delta m^2, \sin^2\theta_{12},
\sin^2\theta_{13}, \sin^2\theta_{23}\right\} \,.
\label{eq:chi2}
\end{align}
Here, $\mathcal{O}_i^{\rm exp}$ stands for the respective
best-fit values and $\Delta\mathcal{O}_i^{\rm exp}$ for the respective 
widths of the full $\pm3\,\sigma$ ranges divided by six, see Tab.~\ref{tab:data}.
For each choice of values for the five experimental
input observables, $\left\{\mathcal{O}_i\right\}$, Eq.~\eqref{eq:chi2}
provides us with a $\chi^2$ value that we can use as a likelihood measure.
Under the assumption of Gaussian distributions and given the fact
that we are working with, in total, five independent degrees of freedom,
$\chi^2$ values of $\chi^2 = \left(5,89, 11.31, 18.21\right)$ correspond
to deviations from the global best-fit configuration at the
$\left(1\sigma,2\sigma,3\sigma\right)$ confidence level, respectively.

\begin{figure}
\begin{center}
\includegraphics[width=0.49\textwidth]{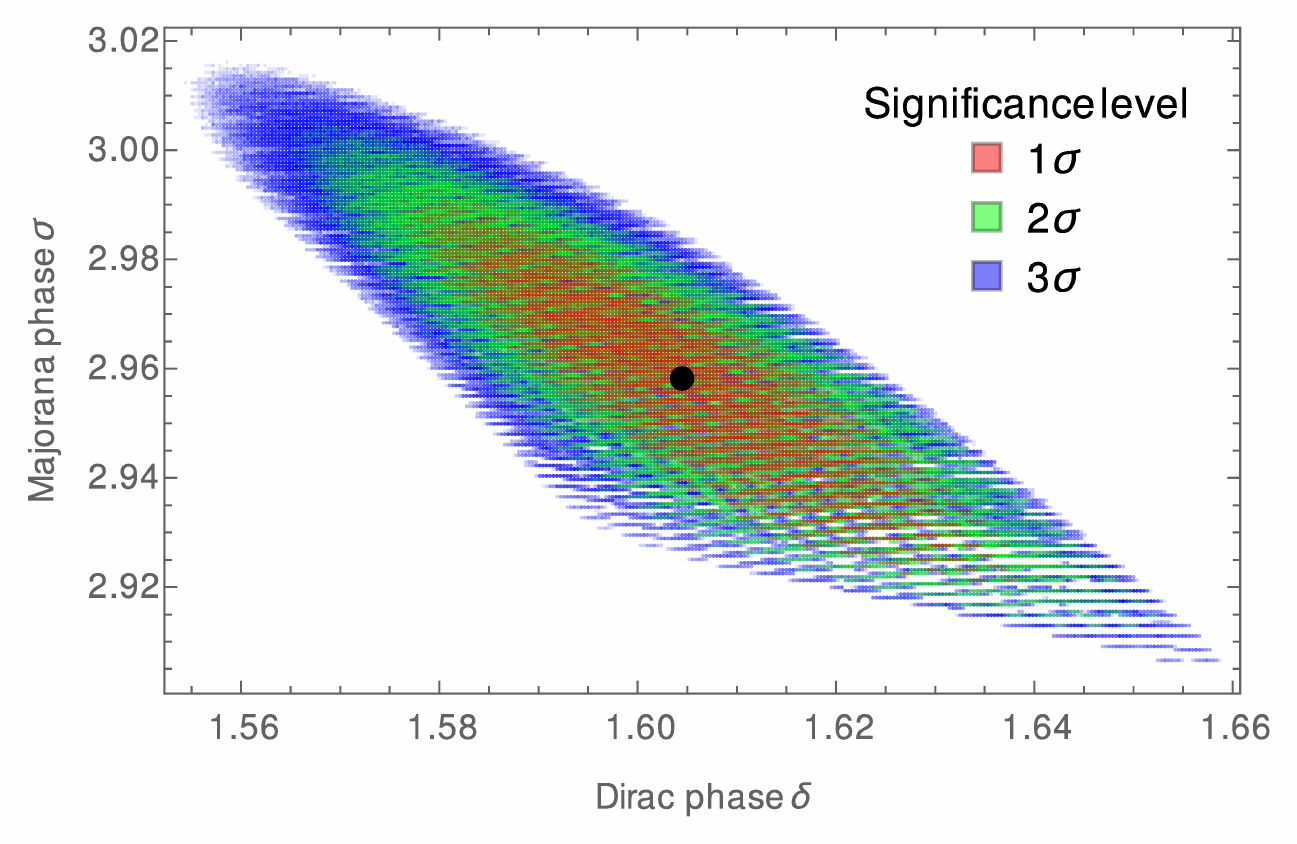}\hfill
\includegraphics[width=0.49\textwidth]{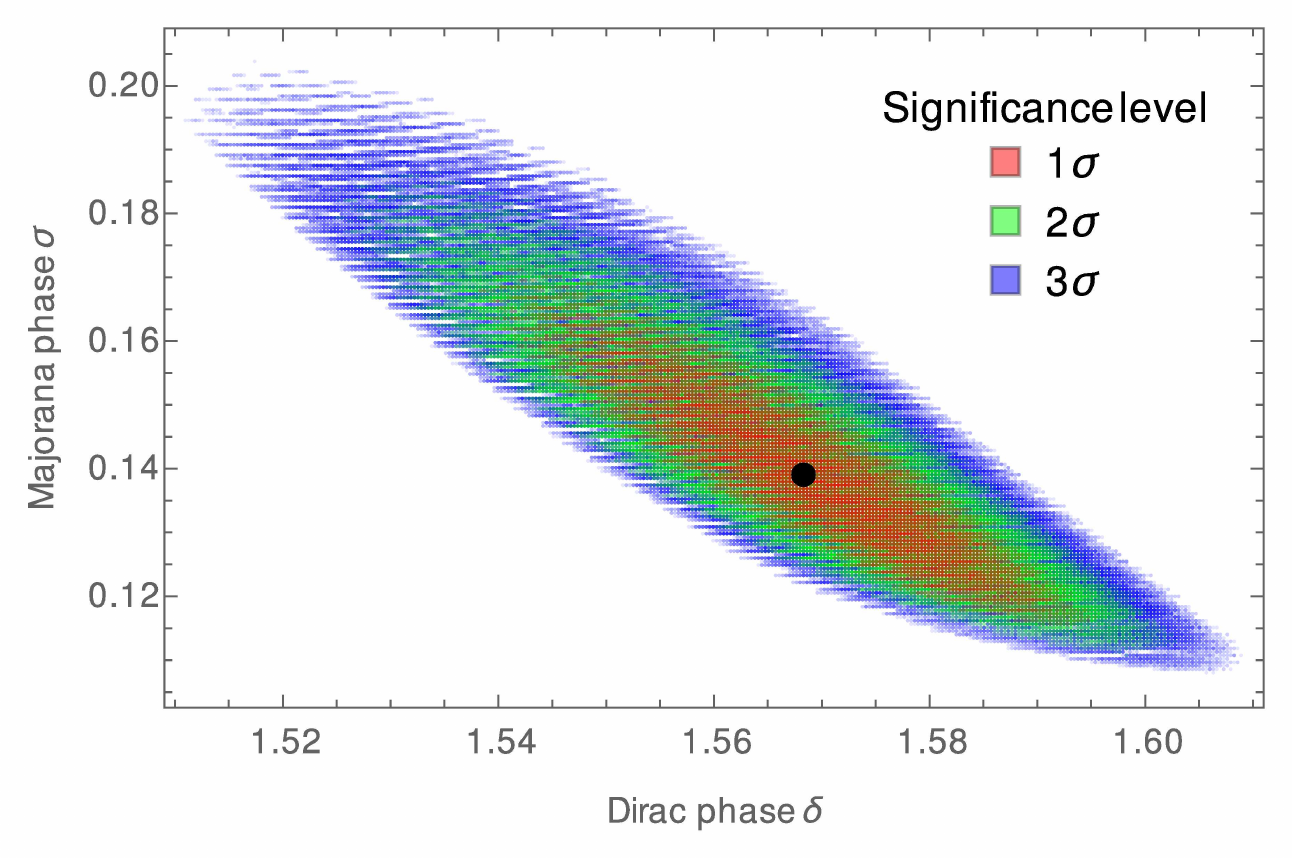}
\caption{Stability of the predictions for $\left(\delta,\sigma\right)$ under variation
of the experimental input parameters $\delta m^2$, $\Delta m^2$,
$\sin^2\theta_{12}$, $\sin^2\theta_{13}$, and $\sin^2\theta_{23}$
for an exact $B_{1,4}$ texture \textbf{(left panel)}
and an exact $B_{2,5}$ texture \textbf{(right panel)}.
The black dots mark the predictions for $\left(\delta,\sigma\right)$ if all observables
are set to their best-fit values, $\left(\delta,\sigma\right) = \left(0.51,0.94\right)\pi$
in the $B_{1,4}$ case and $\left(\delta,\sigma\right) = \left(0.50,0.04\right)\pi$
in the $B_{2,5}$ case, see Eq.~\eqref{eq:dsB1245}.
$\chi^2$ is calculated according to Eq.~\eqref{eq:chi2}.}
\label{fig:expuncert}
\end{center}
\end{figure}

To give an example of how varying the experimental input data
affects our results, we are now going to study the stability of the predictions
in Eq.~\eqref{eq:dsB1245}. 
We proceed in three steps:
(i)~First of all, we vary all five observables within their
experimental $\pm3\,\sigma$ ranges;
(ii)~for each set of input values, we then determine the predictions for
$\delta$ and $\sigma$ according to Eq.~\eqref{eq:T2zero}, assuming
either an exact $B_{1,4}$ or an exact $B_{2,5}$ texture; and 
(iii)~finally, we associate these predictions with the corresponding value of $\chi^2$,
calculated according to Eq.~\eqref{eq:chi2}.
This procedure yields $\chi^2$ as a function of 
$\delta$ and $\sigma$, for both an exact $B_{1,4}$ and an exact $B_{2,5}$
texture, which can, in turn, be translated into $1\,\sigma$,
$2\,\sigma$, and $3\,\sigma$ confidence intervals in the
$\left(\delta,\sigma\right)$ plane.
The outcome of this analysis is shown in Fig.~\ref{fig:expuncert}.
As evident from this figure, the experimental uncertainty
in the neutrino observables induces shifts 
in $\delta$ and $\sigma$ at the level of
$\mathcal{O}\left(1\,\%\right)\pi$---and is, hence, nearly
negligible. 
The requirement of exact texture zeros in the Yukawa matrix
in combination with the current experimental errors determines 
$\delta$ and $\sigma$ to high precision.
This leads us to conclude that all of our results in Sec.~\ref{sec:scan}
ought to be stable under the variation of the input data within
the current experimental limits.
Any further quantitative statement beyond this qualitative conclusion
would require a more sophisticated analysis, which is beyond the scope 
of this work.


\section{Conclusions}
\label{sec:conclusions}


The type-I seesaw mechanism with only two right-handed neutrinos is a minimal
neutrino mass model that manages to account for all observed mass-squared differences
and mixing angles in the SM neutrino sector.
Compared to the conventional three-right-handed-neutrino scenario,
the minimal seesaw model comes with a smaller number of
parameters that are inaccessible at low energies and is, thus, more predictive.
In addition, it complies with the philosophy of Occam's razor,
which calls on us to always only make as few assumptions as necessary.


\begin{figure}[t]
\begin{center}
\includegraphics[width=\textwidth]{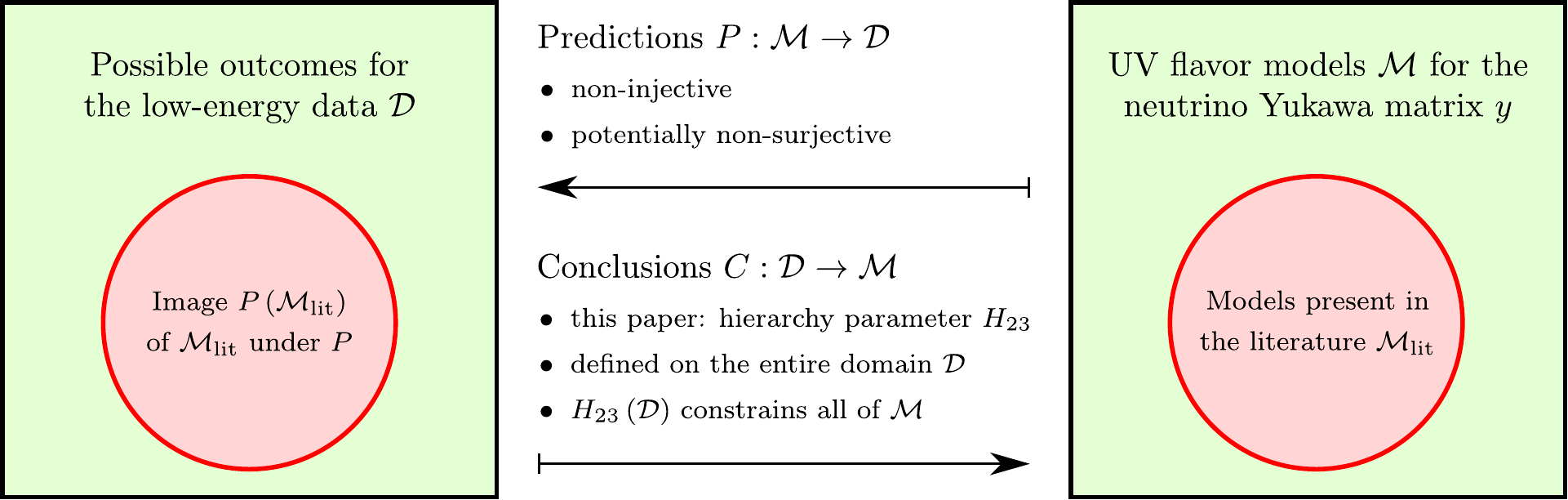}
\caption{Complementarity between conventional model building and our approach in this paper.
Top-down model building \textbf{(right to left)} typically makes non-injective predictions
for the outcome of experimental measurements.
Depending on the specific assumptions made when building UV flavor models,
one may not have a fitting model at hand for each conceivable experimental result.
Meanwhile, it is clear that the list of flavor models in the literature
is not exhaustive and that there are still many models that await
to be worked out.
In our bottom-up approach \textbf{(left to right)}, we remain, by contrast, agnostic
as to what UV dynamics are exactly responsible for the form of the neutrino Yukawa matrix.
By scanning the entire complex $z$ plane, we make sure to cover
all UV completions of the minimal seesaw model, independently of
whether each of these models (i.e., each of these points in the $z$ plane)
has already been recognized as a particularly well motivated flavor structure or not.
This provides us with a function, $H_{23}$, defined on the entire domain
of conceivable experimental data (see Eq.~\eqref{eq:HierParam}).
No matter what values for $\delta$ and $\sigma$ future experiments will find,
once we interpret the new data in the context of the minimal seesaw model,
we will always learn something new about the underlying flavor structure
at high energies.}
\label{fig:maps}
\end{center}
\end{figure}


In the context of this model, we have addressed the following question is
this paper:
Suppose leptonic $CP$ violation should be conclusively measured in the near
future, what lessons could we possibly learn from such a measurement about
the underlying Yukawa structure of the SM neutrino sector?
In particular, we were interested in studying to what extent one would still
be able to realize (approximate) textures zeros in the neutrino Yukawa matrix
for given values of the $CP$-violating phases $\delta$ and $\sigma$.
In order to answer this question, we introduced a novel \textit{hierarchy
parameter}, $R_{23}$ (see Eq.~\eqref{eq:R23}), in terms of a ratio
of ordered Yukawa couplings. 
Minimizing this hierarchy parameter over all ``unphysical''
parameters (i.e., all parameters that are inaccessible at low energies),
we were then able to identify the respective maximal Yukawa
hierarchies for each pair of hypothetical $\left(\delta,\sigma\right)$
values.
This analysis led us to several interesting results:
\begin{itemize}
\item We recovered the exact two-zero textures $B_{1,4}$ and $B_{2,5}$ that can
be realized in the IH case, see Figs.~\ref{fig:H23Scan} and \ref{fig:IHsep}.
In addition, we were able to quantify to what an extent the assumption
of exact texture zeros would need to be relaxed, if the phases $\delta$ and $\sigma$
should eventually be measured at values slightly off the expected values
in these two scenarios, see Fig.~\ref{fig:zoom}.
Besides that, we stress on a more theoretical note that one can imagine many
possibilities how, in the ultimate theory of flavor, texture zeros
may end up being disturbed (e.g., via quantum or gravitational effects).
Against this backdrop, our results in Fig.~\ref{fig:zoom} should be regarded
as the theoretical error bars belonging to the precise predictions in Eq.~\eqref{eq:dsB1245}.
\item We found that, allowing for perturbations of $\mathcal{O}\left(10\,\%\right)$,
approximate two-zero textures may also be realized in the NH case;
e.g., an approximate $A_1$ texture at
$\left(\delta,\sigma\right) = \left(0.55,0.96\right)\pi$ or at
$\left(\delta,\sigma\right) = \left(1.45,0.04\right)\pi$,
see Figs.~\ref{fig:H23Scan} and \ref{fig:NHsep}.
This conclusion represents an important caveat to the conventional wisdom
that the type-I seesaw mechanism involving only two right-handed neutrinos actually
does not allow for two-zero textures in the NH case.
\item We observed that the minimized hierarchy parameter
$H_{23}$ (see Eq.~\eqref{eq:HierParam}) tends to take particularly small
values at $\left(\delta,\sigma\right)$ values that conserve $CP$.
Our list of approximate two-zero textures in Tab.~\ref{tab:results}, therefore,
splits into $CP$-preserving as well as $CP$-violating scenarios.
\item We studied the interplay of ``flavor-aligned'' and ``flavor-nonaligned''
Yukawa matrices, see Figs.~\ref{fig:NHsep} and \ref{fig:IHsep}, arriving at
the conclusion that the possibility
of flavor alignment should not be neglected when searching
for the \textit{global} minima of the hierarchy parameter $R_{23}$.
\item Finally, we checked the stability of our numerical results under
variations of the experimental input data. 
We found that the experimental uncertainties in the low-energy
observables shift our predictions for $\delta$ and $\sigma$ at
the level of $\mathcal{O}\left(1\,\%\right)\pi$, see Fig.~\ref{fig:expuncert}.
This is safely negligible.
\end{itemize}


In closing, we emphasize that we performed a data-driven bottom-up
analysis, see Fig.~\ref{fig:maps}.
Instead of starting from the high-energy perspective (i.e., choosing a particular
UV flavor model and calculating the corresponding predictions for the low-energy
observables), we based our analysis on the question as to what might be seen
in neutrino experiments in the near future.
This is a complementary approach, which allows to derive data-driven
and model-independent constraints on all conceivable UV completions of the minimal
seesaw model, irrespectively of whether these models can already be found in the literature
or still need to be discovered.


Of course, all of our results heavily rely on the assumption of only two
right-handed neutrinos and should only be interpreted in this context. 
If experiments should find that the lightest neutrino mass eigenstate
has nonzero mass, $\textrm{min}\left\{m_i\right\} > 0$, our numerical results
would become obsolete.
Our general strategy, on the other hand, appears to be useful also in the context of other models
and with regard to other types of Yukawa structures.
In the usual seesaw model with three right-handed neutrinos,
one might, e.g., introduce a hierarchy parameter
$R_{mn} = \left|\hat{\kappa}_m\right|/\left|\hat{\kappa}_n\right|$
as a measure for the gap between the $m^{\rm th}$ and the $n^{\rm th}$ entry
in the list of ordered Yukawa couplings $\hat{\kappa}$ (see also Eq.~\eqref{eq:R23}).
Minimizing this parameter over the three complex phases in
the Casas-Ibarra parametrization of the three-neutrino Yukawa matrix,
$\omega_{12}$, $\omega_{13}$, and $\omega_{23}$,
\begin{align}
H_{mn}\left(\delta,\sigma\right) = \min_{\omega_{12}}\min_{\omega_{13}}\min_{\omega_{23}}
R_{mn}\left(\delta,\sigma;\omega_{12},\omega_{13},\omega_{23}\right) \,,
\end{align}
would then allow
to study arbitrary textures in a very general context.
Of course, the technical complexity of such an analysis would
increase with the number of undetermined parameters.
But we believe that, in principle, our method should be applicable
in a broad class of models and for a large variety of Yukawa structures.
In the end, extending our analysis to other neutrino mass models
would serve as an important preparation for the anticipated data
on $CP$ violation in the lepton sector.
Only if top-down model building and bottom-up data analysis go hand in hand,
we will be able to make progress in our understanding of the SM flavor structure.


\subsubsection*{Acknowledgements}


T.\,R.\ would like to thank Moritz Platscher for helpful discussions.
K.\,S.\ would like to thank Tsutomu T.\ Yanagida for
inspiring comments at the early stages of this project.



\end{document}